\documentclass{aa}  

\usepackage{graphicx}
\usepackage{txfonts}

\usepackage{color}

\usepackage{hyperref}
\usepackage[]{natbib}
\bibpunct{(}{)}{;}{a}{}{,} 

\makeatletter
\renewcommand*\aa@pageof{, page \thepage{} of \pageref*{LastPage}}
\makeatother

\begin{document}
    \title{Open cluster dissolution rate and the initial cluster mass function in the solar neighbourhood}
    \subtitle{Modelling the age and mass distributions of clusters observed by Gaia}

   \author{Duarte Almeida\inst{1,2}       
        \and
        André Moitinho\inst{1,2}
        \and
        Sandro Moreira\inst{1,2}
          }
    
   \institute{CENTRA, Faculdade de Ci\^{e}ncias, Universidade de Lisboa, Ed. C8, Campo Grande, 1749-016 Lisboa, Portugal \and Laborat\'{o}rio de Instrumenta\c{c}\~{a}o e F\'{\i}sica Experimental de Part\'{\i}culas (LIP), Av. Prof. Gama Pinto 2, 1649-003 Lisboa, Portugal\\
              \email{duarte.almeida@sim.ul.pt, andre@sim.ul.pt, sandro@sim.ul.pt}
            }

   \date{Received date / Accepted date}

\abstract
{The dissolution rate of open clusters (OCs) and integration of their stars into the Milky Way's field population has been previously explored using their age distribution. With the advent of the \textit{Gaia} mission, we have an exceptional opportunity to revisit and enhance these studies with ages and masses from high quality data.} 
{To build a comprehensive \textit{Gaia}-based OC mass catalogue which, combined with the age distribution, allows a deeper investigation of the disruption experienced by OCs within the solar neighbourhood.}
{Masses were determined by comparing luminosity distributions to theoretical luminosity functions. The limiting and core radii of the clusters were obtained by fitting the King function to their observed density profiles. We examined the disruption process through simulations of the build-up and mass evolution of a population of OCs which were compared to the observed mass and age distributions.} 
{Our analysis yielded an OC mass distribution with a peak at $log(M)$ = 2.7 dex ($\sim 500 M_{\odot}$), as well as radii for 1724 OCs. Our simulations showed that using a power-law Initial Cluster Mass Function (ICMF) no parameters were able to reproduce the observed mass distribution. Moreover, we find that a skew log-normal ICMF provides a good match to the observations and that the disruption time of a $10^4 M{_\odot}$ OC is $t_4^{\text{tot}} = 2.9 \pm 0.4$ Gyr.}
{Our results indicate that the OC disruption time $t_4^{\text{tot}}$ is about twice longer than previous estimates based solely on OC age distributions. We find that the shape of the ICMF for bound OCs differs from that of embedded clusters, which could imply a low typical star formation efficiency of $\leq 20\%$ in OCs. Our results also suggest a lower limit of $\sim 60 M{_\odot}$ for bound OCs in the solar neighbourhood.}

   \keywords{Galaxy: evolution -
                Galaxy: stellar content -
                open clusters and associations: general -
                Surveys: Gaia
               }

   \maketitle
%
\defcitealias{anderson_2023}{AMD23}

\section{Introduction \label{ch:intro}}

Open clusters (OCs) are gravitationally bound stellar systems, typically comprising tens to hundreds of stars that formed together \citep{lada_embedded_2003}. As time progresses, they gradually disperse into the general field star population. The dispersion of OCs is influenced by a combination of internal factors, including the cluster's mass and dynamics, as well as external factors such as galactic tidal forces, interactions with giant molecular clouds (GMCs), and the influence of spiral arms \citep{lamers_clusters_2006}.

Mass plays a crucial role in determining the internal dynamics of star clusters and their interactions with external galactic gravitational forces.  Because a large fraction of stars, possibly most stars, form in clusters \citep[e.g.][]{lada_embedded_2003}, the process of dissolution is instrumental in shaping both the mass and age distribution of the star cluster population we observe, as well as the characteristics of the field star population. Consequently, accurately determining the mass and age distribution of the Milky Way's star clusters is essential for refining models of their evolution and enhancing our understanding of the dissolution processes they undergo.

Prior to the \textit{Gaia} mission \citep{gaia_mission_ref} of the European Space Agency (ESA), OC masses had only been published for about 35\% of the known clusters in the Milky Way.  The majority of this data was provided by \citet{piskunov_initial_2008} using photometry and astrometry compiled in the ASCC$\sim$2.5 catalogue \citep{kharchenko_all-sky_2001}, which is limited to stars brighter than approximately V$\sim$12 mag. This calls for revisiting OC masses using the exquisite data delivered by \textit{Gaia}, whose various data releases \citep{2016A&A...595A...2G, gaia_2_2018, 2021A&A...649A...1G} have provided astrometric and photometric data for nearly two billion stars, reaching  $G \sim$ 20.5 mag, with photometric and astrometric uncertainties at the mmag and sub-milliarcsec levels, respectively. 
This has brought much improved lists of OC members, as well as an avalanche of discoveries of new OCs and candidates \citep[e.g.][]{castro_2018_nearby, liu_catalog_2019, cantat_gaudin_incomplete_2019, sim_207_2019, castro-ginard_hunting_2020, ferreira_new_2021, Hunt_I_2021}, and large scale determinations of OC distances, ages, extinctions and other parameters \citep[e.g.][]{cantat-gaudin_gaia_2018, bossini_age_2019, cantat-gaudin_painting_2020, monteiro_fundamental_2020, dias_updated_2021, Hunt_II_2023} published. 

However, large scale \textit{Gaia}-based determinations of OC masses have been missing until very recently. \citet*{anderson_2023} \citepalias[hereafter][]{anderson_2023} published a catalogue of masses for 773 OCs.  \citet{hunt_III_masses} published a catalogue of OC masses for 3530 clusters, with they estimated the age and mass distributions in the Milky Way. On a smaller scale, \citet{Cordoni_2023} determined masses for 78 OCs in their study on the role of binaries in OC evolution. The work presented in this article has different aims and follows a different approach for calculating masses, as discussed in Sect.~\ref{ch:mass_det}.

In this work, we aim to determine the timescale of the disruption experienced by clusters in the solar neighbourhood. What differentiates this study from previous works \citep[e.g.][]{lamers_analytical_2005,anders_star_2021} is the incorporation of both the age and mass distributions in our analysis, as well as the use of \textit{Gaia} data. Estimates for the disruption parameters have been previously found using the age distribution and N-body simulations of clusters in the tidal field of the Galaxy. Considering that the cluster disruption time depends on the initial mass M of the cluster as $t_{dis} = t_0(M/M_\odot)^{\gamma}$, \citet{boutloukos_star_2003} found $\gamma$ = 0.6 for clusters within 1 kpc of the Sun, while \citet{baumgardt_dynamical_2003} obtained $\gamma$ = 0.62 from N-body simulations of clusters in the tidal field of our galaxy at different galactocentric distances. \citet{lamers_analytical_2005} compared the predicted to the observed age distribution of open clusters in the solar neighbourhood, from the \citet{Kharchenko_2005} catalogue, and obtained $t_0$ = $3.3^{+1.5}_{-1.0}$ Myr and $\gamma$ = 0.62. However, the previous studies focused solely on the age distribution of clusters due to the lack of a large-scale mass catalogue so, in this work, we aim to expand our understanding by using both the age and mass distributions. 

To achieve this, we build a catalogue of open cluster luminous masses in the Milky Way, using data from the \citet{dias_updated_2021} catalogue and \textit{Gaia} DR2 data. Determining the mass bound to clusters requires also determining their limiting radii. Then, we generate simulated distributions of age and mass, which are compared to the observations of clusters located within a 2 kpc radius around the Sun, with ages under 1 Gyr. By comparing the simulated and observed distributions for various parameter values, we find the disruption parameters that best match the observations.

This paper is organized as follows. Chapter \ref{ch:data} describes the observational data used in this work. Chapters \ref{ch:methods_general} and \ref{ch:results} present the methods and results for the masses and radii of 1724 OCs, followed by the investigation of the OC disruption timescale in the solar neighbourhood (Chapters \ref{ch:disruption} and \ref{ch:disruption_timescale}). Chapter \ref{ch:conclusion} closes this work with the conclusions and final considerations. 

\section{Data \label{ch:data}}

We use the \citet{dias_updated_2021} OC catalogue which has entries for 1743 objects and is based on \textit{Gaia} DR2 \citep{gaia_2_2018}. The listed cluster parameters, namely distance, reddening, age and metallicity were homogeneously derived applying a cross-entropy isochrone fitting method described in \citet{monteiro_opd_2017} and \citet{monteiro_distribution_2021}. 
Additionally, \citet{dias_updated_2021} provide tables of individual stellar membership probabilities. The lists of members were compiled from other published studies. Some of those studies \citep{liu_catalog_2019,castro-ginard_hunting_2020,castro-ginard_hunting_2022} do not provide membership probabilities. In these cases, the probabilities were computed by  \citet{dias_updated_2021}. The membership tables are limited to stars brighter than  $G \sim$ 18 mag.  

We note that, in principle, we could have adopted the catalogue of \citet{cantat-gaudin_painting_2020}. Both catalogues have comparable sizes and parameters and compile similar lists of members from the same sources. However, as discussed above, membership probabilities are not provided by some sources. In such cases, \citet{cantat-gaudin_painting_2020} assigned a value of 1. Since our analysis uses individual membership probabilities, we have opted for using \citet{dias_updated_2021}.

We also acknowledge the recent catalogues of OCs published by \citet{Hunt_II_2023, hunt_III_masses}, which include many new entries with computed ages, distances, and radii. As these catalogues were published during the final stages of our work, they were not included in our baseline compilation of OCs. Nevertheless, comparisons with these catalogues are provided in Sect.~\ref{ch:compare_radius}.

\section{Methods \label{ch:methods_general}}

One of the main aims of this study is the determination of OC masses, i.e. the total mass of stars bound to a cluster. However, cluster masses can be determined from different observables. Using mass-luminosity relations, we may derive cluster masses from stellar luminosity measurements. These mass determinations are often referred to as "photometric" or "luminous masses" and are the ones determined here. Masses derived from kinematic measurements are referred to as "dynamical masses" and "virial masses" (for clusters assumed to be in virial equilibrium). Additionally, one may also consider "tidal masses" which are masses deduced from a cluster's limiting radius, which is imposed by the balance between the cluster self gravity and the galactic tidal forces exerted upon the cluster \citep[e.g.][]{piskunov_tidal_2008}.  

\subsection{Member selection}

To produce cluster member lists as complete as possible, while simultaneously minimize the contamination by field stars, as a general rule, we selected stars with membership probabilities above 50\%. However, when analysing their colour-magnitude diagrams (CMDs), some clusters still showed clear  contamination. In those cases, we increased the membership cut-off to reduce as much as possible the contamination while conserving the clear cluster members. Some clusters had poorly defined CMDs (even when relaxing membership cut-offs). We have thus visually inspected the CMDs of all 1743 OCs, with PARSEC isochrones \citep{2012MNRAS.427..127B} fits over-plotted, attributing a classification, P1 (good), P2 (medium) and P3 (worst), based on the dispersion of the cluster sequence and on the quality of the isochrone match. As shown in Table~\ref{tab:P_class}, the majority of our sample has classification P1. Additionally, as we will later compare the masses derived using different \textit{Gaia} bands, only stars detected in the 3 bands (G, $G_{BP}$ and $G_{RP}$) were considered.

\begin{table}[ht]
\caption{Number of open clusters per classification.}
\centering
\begin{tabular}{lll}
\hline\hline
Class & Number & Fraction \\ \hline
P1 & 931 & 54.0 $\%$ \\ 
P2 & 641 & 37.2 $\%$ \\ 
P3 & 152 & 8.8 $\%$ \\ \hline
\end{tabular}
\label{tab:P_class}
\end{table}

Stars outside the cluster tidal radius are dispersing into the field population and can be found distributed as diffuse tails and coronae around clusters \citep[e.g.][]{2001A&A...377..462B,2015MNRAS.449.1811D,2021A&A...645A..84M,2023A&A...673A.128G,2023arXiv231202263D}. They have kinematic and photometric distributions close to those bound to the cluster. Therefore, separating the bound and unbound populations requires additional criteria. Here we will make the approximation that all stars within the cluster radius contribute to the bound mass. As will be seen in Sect.~\ref{ch:compare_radius}, within the uncertainties, this assumption should not significantly affect our results. 

Therefore, the next step towards building the catalogue of OC masses is to determine their radii.

\subsection{Radii determination \label{ch:radii_det}}

For the radii determinations, we fitted the King empirical profile \citep{king_structure_1962} to the OC radial density profiles (RDPs). To account for the contamination of foreground and background stars, we follow \citet{kupper_peculiarities_2010} and introduce an additional parameter $c$. 
The modified King function provides the number of stars per $pc^2$ in the form of:
\begin{equation}
    n(R) = \begin{cases} N_0 \left( \frac{1}{\sqrt{1+(R/R_c)^2}} - \frac{1}{\sqrt{1+(R_k/R_c)^2}} \right)^2 + c & if \ R < R_k \\ c & if \ R \geq R_k  \end{cases}
\end{equation}
Where $R_k$ is the limiting radius, beyond which the density of stars is indistinguishable from the density of the background. $R_c$ is the core radius, which is the radius for which the density fall to half of the central density. $N_0$ is a scaling factor which reflects the central density of the cluster. \textit{c} is the background density, taken as constant.

For simplicity, we assume that the limiting radius derived from the King profile is equivalent to the tidal radius at which the gravitational attraction of the cluster balances the external tidal forces. However, it is important to acknowledge that this approximation holds true only when the stars within the cluster entirely fill the cluster's Jacobi radius. This condition may not always be fulfilled, and thus, we discuss the effect of the radius on the derived mass in Sect. \ref{ch:mass_error_analysis}. To make the difference clear, we represent the limiting radius as $R_k$ (King radius) instead of $R_t$ (tidal radius) as usually found in the literature.

The ring width for the RDPs was chosen as 0.5 pc for clusters where the farthest star was less than 12 pc from the centre, and 1 pc for the other cases, with the intention of avoiding under or over sampling the density profile in clusters with fewer stars.

The profile fits were done with the LMFIT Non-Linear Least-Squares Fitting python package \citep{newville_matthew_2014_11813}.  We used the Maximum Likelihood Estimation (MLE) from the \textit{emcee} method within LMFIT to obtain the King parameters. The uncertainties of each King parameter were taken to be $\pm \sigma$, as provided by the output of \texttt{LMFIT}. These uncertainties are sampled from the posterior distribution, marginalised over the other King model parameters. The convergence of the chains was evaluated based on the value of the integrated autocorrelation time ($\tau$) which is related to the Monte Carlo error (or sampling error), as recommended by \citet{goodman_2010}. In cases where the program raised a warning due to the value of $\tau$, the number of steps was gradually increased (starting from 100,000) until the criterion of $\tau$ > 50 was satisfied. 

To ensure the results were physically meaningful, the King radius was forced to be larger than the core radius and was allowed to vary between 0.5 and 100 pc. The core radius varied between 0.2 pc and the distance of the farthest star to the centre of the cluster.

\subsection{Mass determination \label{ch:mass_det}}

The luminous mass of each open cluster was determined by comparing the observed luminosity distribution to the theoretical luminosity function (LF) for each cluster. The luminosity functions were obtained from the web interface for the PARSEC\footnote{\url{http://stev.oapd.inaf.it/cgi-bin/cmd}} models of stellar evolution \citep{2012MNRAS.427..127B,marigo,marigo_scale_rel,chen_2015} and were calculated for the \textit{Gaia} filter passbands of \citet{mariz_weiler_2018}, considering the \citet{kroupa_2001} initial mass function corrected for unresolved binaries of stars, and solar metallicity $Z = 0.0152$.

Our mass determination method consists in fitting (scaling) the model LF to the observed LF. We start by selecting the model LF with the same age as the cluster being fit, and add to it the cluster's distance modulus and interstellar absorption. We then scale the model LF to the observed LF. The scaling factor is the number by which the model LF is multiplied that minimizes, in a least squares sense, the deviation to the observed LF. 

It must be noted that the PARSEC luminosity functions are provided normalized, corresponding to a conceptual population born with 1$M_{\sun}$. However, at any given age, the LF no longer corresponds to 1 solar mass since some stars will have died or lost mass. This means that the scale factor obtained above gives the cluster mass at birth. Because we are interested in the mass at the present age of a cluster, a scaling correction must be applied.  To obtain this correction, a population of stars with known mass was generated using the PARSEC web interface and the mass of the population was evaluated (by summing the masses of each star) at different ages, t, from $\log(t)$ = 6.6 to 10, with steps of 0.1 dex. We find that for the age range considered, mass decreases linearly with $\log(t)$, yielding a correction  ${\phi}_M = - 0.135 \log(t) + 1.781$, where ${\phi}_M$ is the multiplicative correction to transform birth mass into present mass.

As a sanity check, a second mass estimation method was used. Here we considered the relation between the area under the observed and theoretical LFs. Dividing the areas under both distributions gives the scaling factor between them, which multiplied by the scaling correction (${\phi}_M$) yields the mass of the cluster. This method does not capture the mass (luminosity) and age dependent fine features of the LF, and is thus a less accurate approach.

As the PARSEC web interface only allows to select one bin width in each retrieval, our freedom of choice for the bin width was limited. Choosing a different bin width for every one of the 1743 clusters would become unpractical, so we chose a common bin width of 0.5 mag for most clusters. However, when the highest count bin had fewer than 10 stars, a bin width of 1 mag was used instead. We visually inspected the LF of the 556 clusters in this situation to verify if they were grossly over- or under-sampled.

As previously mentioned, our stellar samples are limited to apparent magnitude $G \leq$ 18 mag, which is well within the Gaia completeness limit. However, when the apparent magnitudes are converted to absolute magnitudes and discretized in bins, the last absolute magnitude bins might not be complete. To account for this effect, we removed the last bin of the LFs. 

\section{Results: radii, masses and sample selection \label{ch:results}}

\subsection{King radii}

During the analysis, we encountered some cases where the fitting of the King profile did not converge. A deeper examination revealed that the density in the peripheral zones of these cluster fields was not zero but was nevertheless very low, of the order of $10^{-2}$ stars/$pc^2$. We set this value as the background density, $c$, and considered that the King radius should not extend into this region.  This cut-off reduced the incidence of non-converging radius fits.

Despite this improvement, there were still 15 clusters for which the fits failed, even after trials with other membership and background density cut-offs. We found that their density profiles were either nearly flat or exhibited irregular contours, rendering them unsuitable for King profile fitting. Consequently, these objects were omitted from our sample.

Additionally, a visual inspection of the stellar distributions on the sky showed that four objects (Berkeley~58, Berkeley~59, Blanco~1 and NGC~7789) had incorrectly listed centres, clearly off the star distributions. These clusters were also removed from the sample. This leaves us with a sample of 1724 OCs.

\begin{figure}[ht]
\centering
\includegraphics[width=0.95\linewidth]{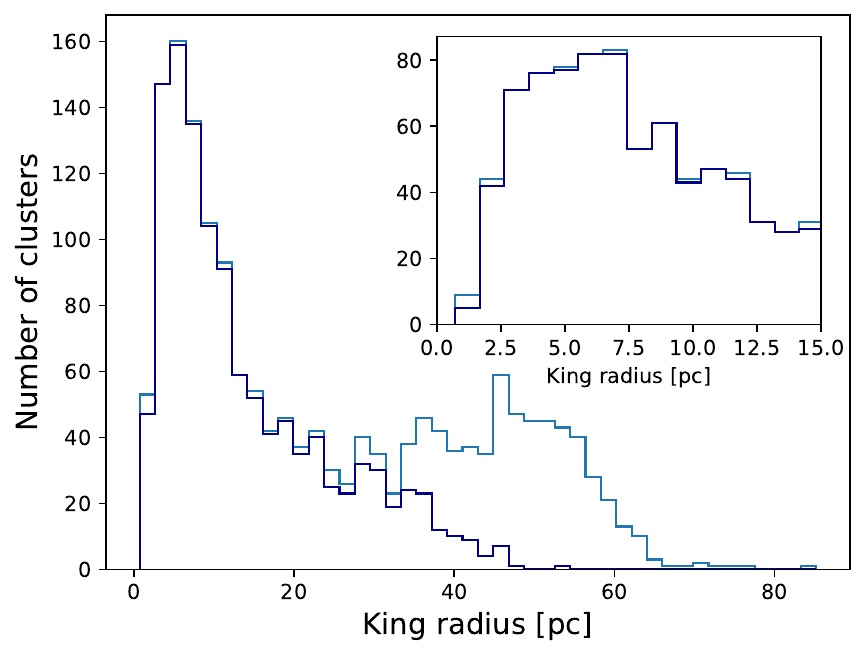}
\caption{Distribution of King radii for 1724 OCs (in light blue). The dark blue histogram represents the distribution of King radii excluding clusters with poor quality fits (classified as R4, see Sect. \ref{ch:raddi_classif}). Bin widths were set using Knuth's rule \citep{knuth_optimal_2006}.}
\label{fig:rk}       
\end{figure}

Fig.~\ref{fig:rk} presents the distribution of King radii of our OC sample. The full dataset, illustrated in light blue, exhibits a wide bump around 50 pc. Such large radii prompted a more detailed examination. To this end, we excluded clusters with poorer fit quality (classified as R4, see Sect. \ref{ch:raddi_classif}) and reevaluated the distribution of radii excluding these clusters, depicted in dark blue. The elimination of these clusters led to the disappearance of the bump. This highlights the importance of careful quality control, which in our case resulted in assigning objects to different quality classes, as will be discussed in Sect.~\ref{ch:raddi_classif}.

The cleaned distribution of radii, excluding object with poor-quality fits, peaks at 5-6 pc, and has a median value of 10 pc, consistent with previous literature values \citep[e.g.][]{piskunov_towards_2007, kharchenko_global_2013}. The distribution of parameters $N_0$ and $c$ as well as the distribution of core radii, with a median value of 2 pc, are shown in Appendices~\ref{ch:appendixA} and~\ref{ch:appendixB}, respectively.

The uncertainties associated with the King and core radii were determined as detailed in Sect. \ref{ch:radii_det}. Our analysis revealed that the median fractional uncertainties for the King radius are 47$\%$ for the lower bound and 95$\%$ for the upper bound. These figures underscore the challenges involved in accurately measuring King radii (defined by the low density regime), especially for clusters with sparse populations. For the core radius (high density regime), as expected the median uncertainties were found to be lower, at 23$\%$ for the lower bound and 25$\%$ for the upper bound.

To assess the robustness of our fitting procedure, we performed a sanity check by iterating over the King radius in increments of 0.5 pc, while allowing the remaining parameters to be optimized by LMFIT for each radius value. For each iteration, we computed the root mean square (RMS) difference between the modelled King profile and the observed stellar density distribution. The parameter set yielding the lowest RMS was selected as the optimal solution. We then compared these results against those obtained through our main method. This comparison revealed that for determinations classified as high quality, the median discrepancy in radii between the two methods was only 2$\%$. However, for the intermediate and low quality cases, the differences increased to 15$\%$ and 37$\%$, respectively. Further analysis confirmed that a significant portion of the clusters contributing to the bump in the distribution of King radii around 50 pc encountered convergence issues during this process, consistently returning in the maximum King radius value preset in this sanity-check loop.

\subsubsection{Radii classification \label{ch:raddi_classif}}

Our procedure yielded radii for a total of 1724 OCs. However, as previously mentioned, not every determination was deemed reliable, particularly in instances where the observed radial density profiles could not be accurately modelled by a King function.

\begin{figure}[ht]
    \centering
    \includegraphics[width=0.95\linewidth]{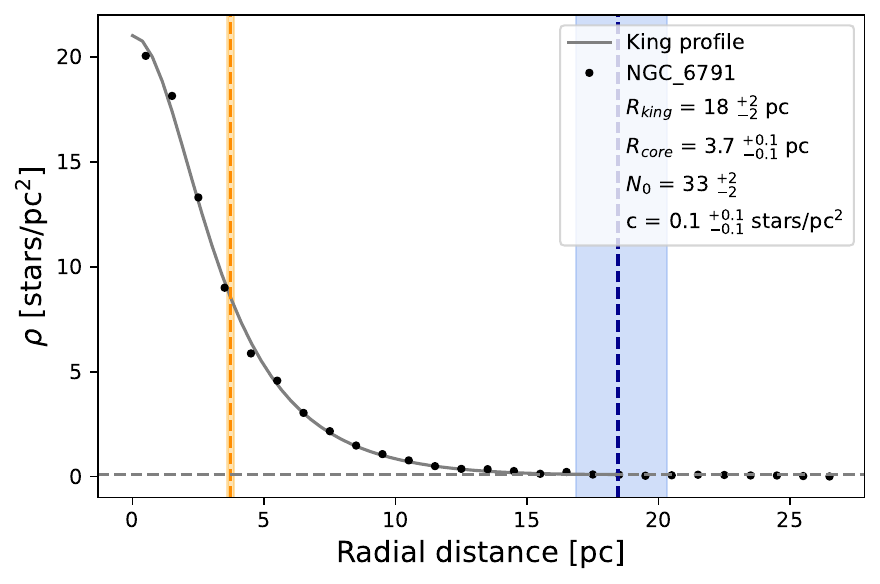}\    
    \includegraphics[width=0.95\linewidth]{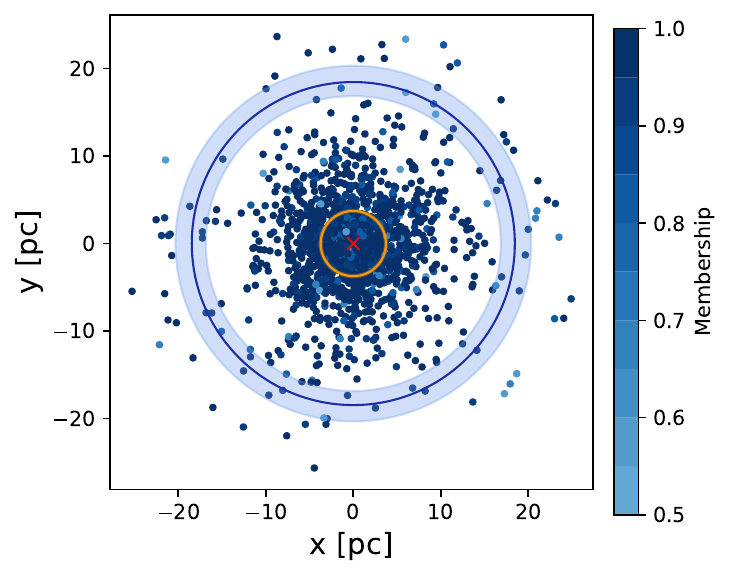}
    \caption{Example of classification R1. (Top) Radial density distribution of open cluster NGC~6791 (black points) with fitted King profile (grey solid line). Blue and orange vertical dashed lines represent the King and core radius, respectively. The upper and lower uncertainty of each radius are displayed as the shaded area around each vertical line. The grey horizontal dashed line represents parameter c, which is the background density. (Bottom) Spatial distribution of the cluster stars onto the tangent plane, colour-coded by membership probability, with blue and orange circles at the King and core radius, respectively.}
    \label{fig:R1_class}
\end{figure}

To address this, we performed a visual classification, assigning each cluster to one of four categories based on the quality of the King profile fit and the congruence of $R_k$ (King radius) and $R_c$ (core radius) with the observed stellar distribution on the tangent plane. These categories are: R1 (best quality), R2 (intermediate quality), R3 (worst quality), and R4 (non-reliable). Specifically, the R3 category includes clusters where the King and core radii are similar, hinting at potential issues with fit convergence. The R4 category primarily includes clusters with excessively large King radii (around 50 pc) or those where the fit is visually unsatisfactory, as well as cases where the density distribution does not have a profile suitable for King function fitting. The distribution of clusters across these classifications is given in Table~\ref{tab:R_class}. Additionally, Fig.~\ref{fig:R1_class} shows a typical example of a cluster classified as R1 (best quality).

\begin{table}[ht]
\caption{Number of open clusters per classification according to the quality of the radii determination.}
\centering
\begin{tabular}{lll}
\hline\hline
Class & Number & Fraction \\ \hline
R1 & 342 & 19.8 $\%$ \\ 
R2 & 495 & 28.7 $\%$ \\ 
R3 & 338 & 19.6 $\%$ \\ 
R4 & 549 & 31.9 $\%$ \\ \hline
\end{tabular}
\label{tab:R_class}
\end{table}

\subsubsection{Comparison with other studies \label{ch:compare_radius}}

We compared the cluster limiting radii obtained in this study with those given in \citet{tarricq_structural_2022}, \citet{Just_2023} and \citet{hunt_III_masses} for which we have 109, 938 and 1264 OCs in common, respectively.

\citet{tarricq_structural_2022} extended member lists to the outskirts of the clusters, including tails and coronae, resulting in significant higher numbers of stars  with respect to this study. The cluster distances employed also differ, with \citet{tarricq_structural_2022} which used the values from \citet{cantat-gaudin_painting_2020} whereas we used \citet{dias_updated_2021}. While both our study and \citet{tarricq_structural_2022} fitted King density profiles, they employed the MCMC sampler \textit{emcee}  \citep{foreman-mackey_emcee_2013} with different criteria and excluded poorly constrained results based on the uncertainties in the radii. In our study, we filtered the results based on the visual quality of the King profile fit. 

The radii reported in \citet{hunt_III_masses} are estimates of the OC Jacobi radii. These were determined by summing the individual stellar masses up to the radius at which the stars can remain bound to the cluster, based on the collective mass of the enclosed stars and the Galactic tidal forces at the cluster's position. For clusters that do not fill their Roche volume, these radii are expected to be larger than those derived from the radial density distribution in our study.

The cluster radii used in \citet{Just_2023} are a compilation from a series of papers \citep{kharchenko_global_2013, Schmeja_2014, Scholz_2015}, based on the Milky Way Star Clusters survey \citep[MWSC][]{kharchenko_all-sky_2001} which has a limiting magnitude of $V\sim 14$ mag. The radii determinations were performed using the cumulative King profile instead of the linear form to mitigate the effects of poor statistics. 

\begin{figure}[ht]
    \centering         
    \includegraphics[width=0.95\linewidth]{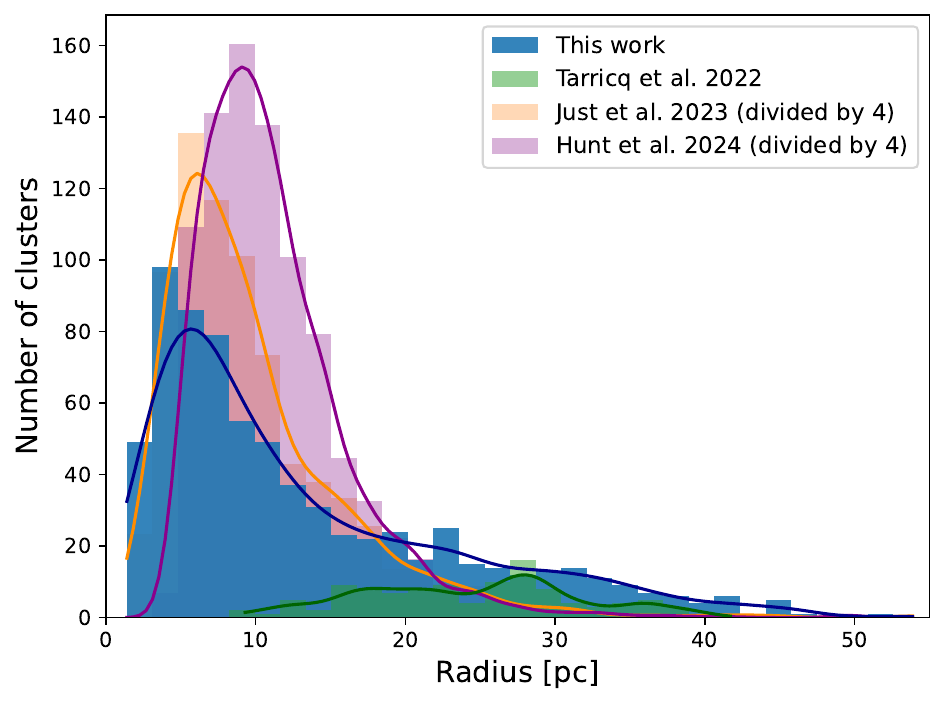}\
    \caption{Radii distributions from this work (blue), \citet{tarricq_structural_2022} (green), \citet{hunt_III_masses} (purple), and \citet{Just_2023} (orange) with KDEs as solid lines. The distributions from \citet{hunt_III_masses} and \citet{Just_2023} were divided by 4 to allow for a comparison as the number of OCs are much higher than in the other catalogues.}
    \label{fig:pisk_tar_rk}
\end{figure}

The radii distributions obtained by \citet{tarricq_structural_2022, hunt_III_masses, Just_2023} and in this study are presented in Fig.~\ref{fig:pisk_tar_rk}. 
This figure displays histograms as well as Kernel Density Estimations (KDE) using the $gaussian\_kde$ function of the Python Scipy \citep{2020SciPy-NMeth} package, with a bandwidth of 2~pc. In general, we note similar distributions of radii, except for those from \citet{tarricq_structural_2022} that tend to be quite higher than the other studies, likely reflecting the effect of using stars in cluster coronae and tails in their radii determinations. 

\begin{figure}[ht]
    \centering   
    \includegraphics[width=0.99\linewidth]{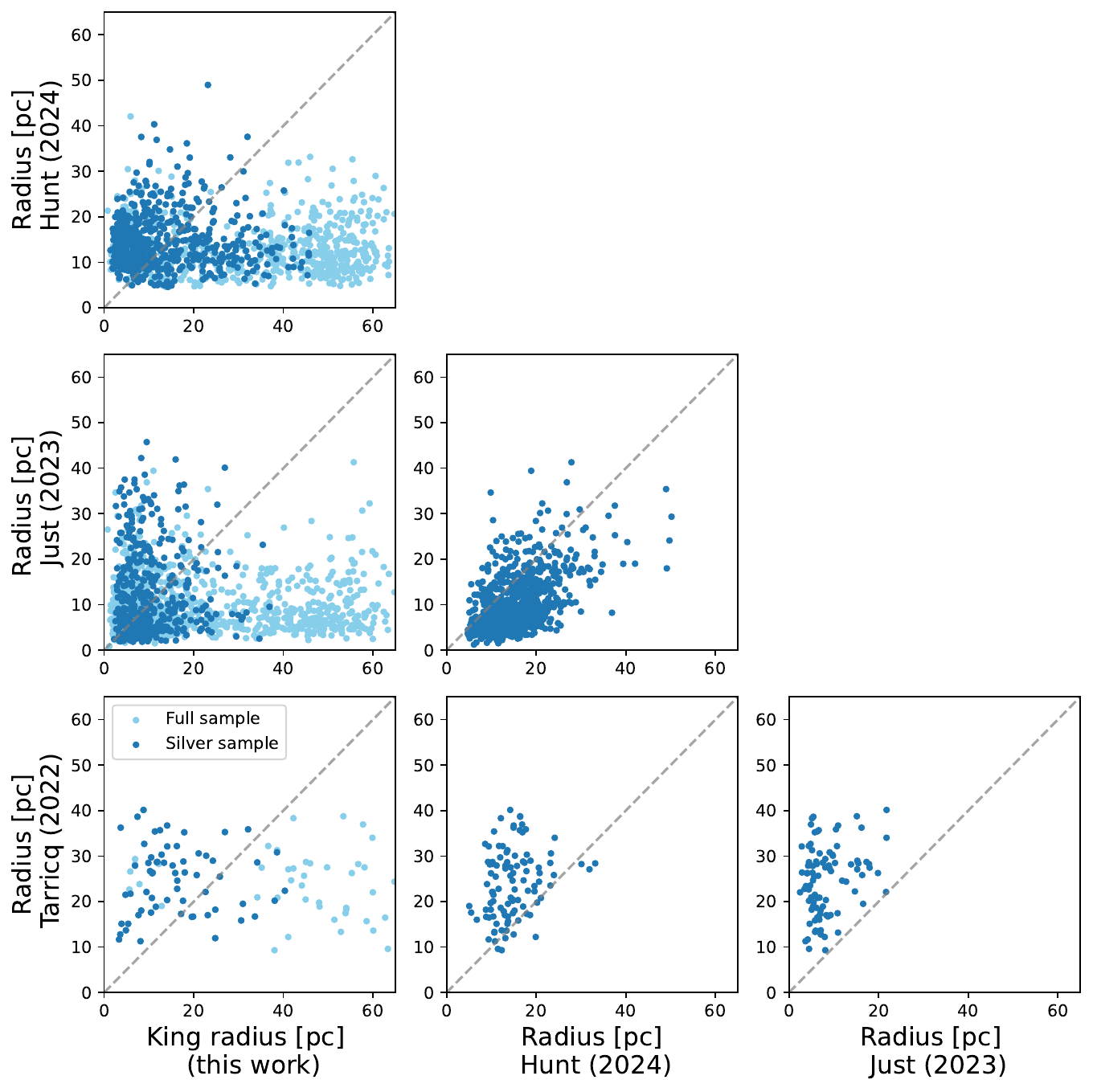}
    \caption{Comparison of the distribution of radii from this work, \citet{tarricq_structural_2022}, \citet{hunt_III_masses} and \citet{Just_2023}.}
    \label{fig:pisk_tar_rk_pts}
\end{figure}

A more enlightening picture emerges in Fig.~\ref{fig:pisk_tar_rk_pts}, where we present the comparisons between the individual radii from each of the catalogues. Here we see that even for catalogues with similar radii distributions the individual values are almost non-correlated! When comparing only with the "silver sample" from this study (our baseline sample, defined in Sect. ~\ref{ch:sample_select}), we find a smaller scatter (mostly from eliminating our higher radii) and that our radii tend to be smaller than the other studies. This might be taken as an indication that the large radii reported in other studies could also be filtered out employing similar quality cuts. We note that errors will lead mostly to larger radii because they are produced by the low signal (low density) regions. Higher density regions produce clearer cluster signals that are in general not confused with the background field level.  In any case, it is clear that the agreement in the distributions does not reflect agreement of the individual values of radii between different catalogues. The large observed scatter highlights the difficulties in determining cluster limiting radii despite the careful analyses done in these studies, and shows that further investigation is needed. 

As we will show in Sect. \ref{ch:mass_error_analysis}, this is not critical for our study since luminous mass determinations turn out not to be much affected by radii errors. However, limiting radii errors will have critical effect on tidal mass determinations.

\subsection{Luminous mass \label{ch:mass_results}}

\begin{figure}[ht]
    \centering
    \includegraphics[width=0.95\linewidth]{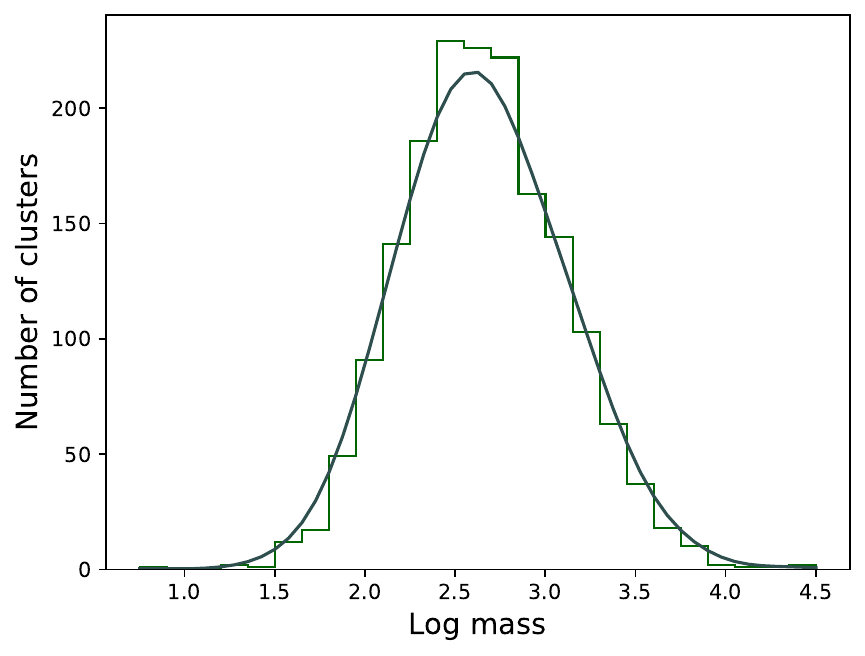}
    \caption{Distribution of the luminous mass, in logarithmic scale. The dark grey solid line is a Gaussian KDE with bandwidth of 0.18 dex.}
    \label{fig:mass_distrib}
\end{figure}

We adopted a conservative approach by using the membership cut-offs determined through visual inspection of the colour-magnitude diagrams as described in Sect.~\ref{ch:methods_general}, which are always above 50$\%$. 
The resulting distribution of mass, from the $G$ band luminosity functions, is shown in Fig. \ref{fig:mass_distrib}. The logarithmic mass distribution is represented with a histogram and a KDE with a bandwidth of 0.18 dex. The KDE shows a peak at $\log(M)$ = 2.7 dex, with a standard deviation of 0.4 dex, which corresponds to a median mass of $\sim$ 500 $M_\odot$.

\subsubsection{Error analysis \label{ch:mass_error_analysis}}

\begin{figure}[ht]
    \centering
    \includegraphics[width=0.95\linewidth]{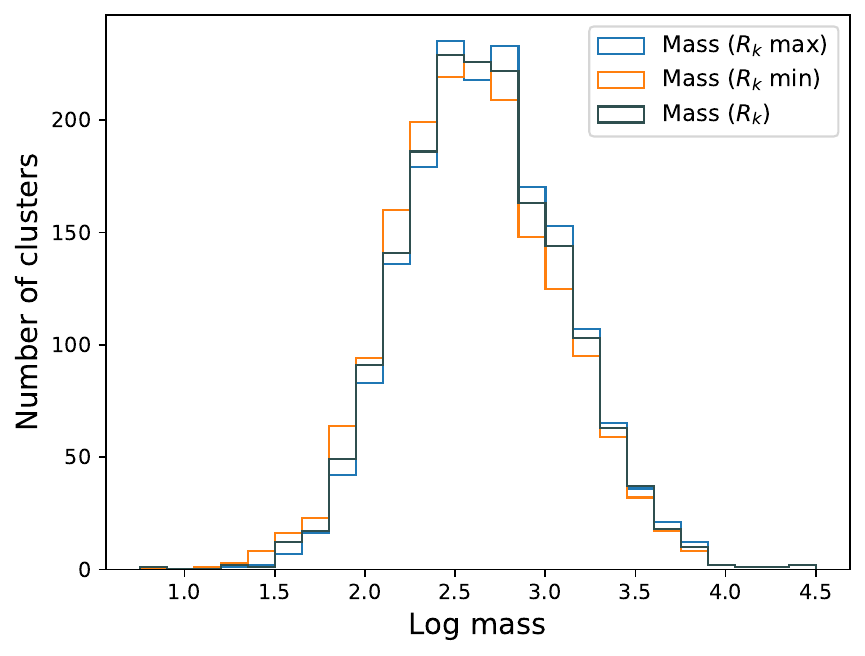}
    \caption{Distribution of the luminous masses determined using only stars inside $R_k$ (grey) and its lower and upper bound (orange and blue, respectively).}
    \label{fig:logmass_min_max}
\end{figure}

In order to investigate the impact of the uncertainties from the King radius on the mass estimates, we computed the mass of each cluster by taking into account the lower and upper bounds of $R_k$. The mass distributions are displayed in Fig. \ref{fig:logmass_min_max}. When considering the lower bound of $R_k$, the mass is decreased by approximately 8\% (median value), whereas for the upper bound it increases by about 6$\%$ compared to the mass within $R_k$. 

\begin{figure}[ht]
    \centering
    \includegraphics[width=0.97\linewidth]{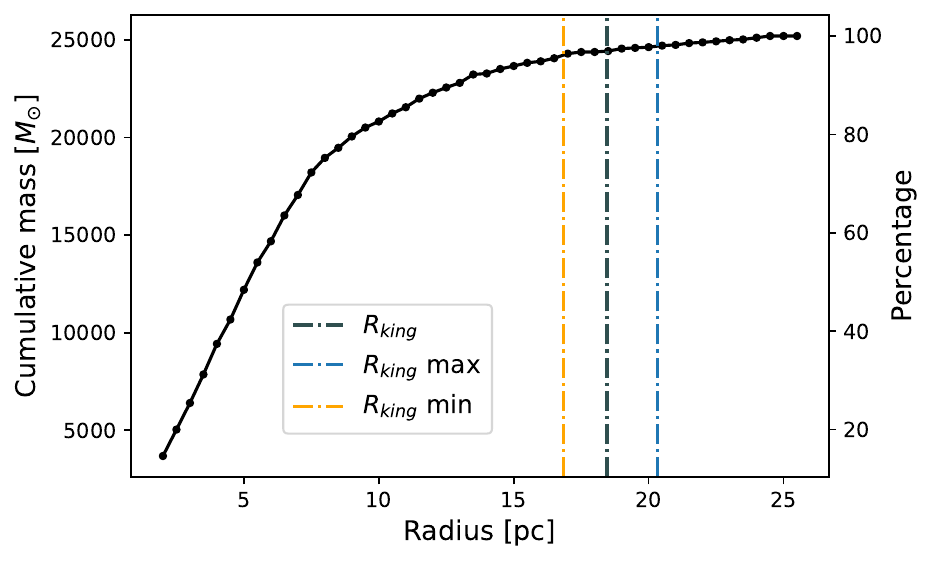}
    \caption{Cumulative mass distribution for open cluster NGC~6791, with vertical dashed-dotted lines at $R_k$ (grey) and its lower and upper bound (orange and blue, respectively).}
    \label{fig:mass_rk_effect}
\end{figure}

As an example, we present the cumulative mass distribution for open cluster NGC~6791 (Fig.~\ref{fig:mass_rk_effect}) with $R_k$ and its lower and upper bounds. As seen, the range of values for the King radius are in the outer regions where less mass is concentrated and the variation of mass is around 1\% in the lower and upper bounds. This illustrates the weak influence of the King radius uncertainties on the luminous mass of the cluster.

\begin{figure}[ht]
    \centering
    \includegraphics[width=0.9\linewidth]{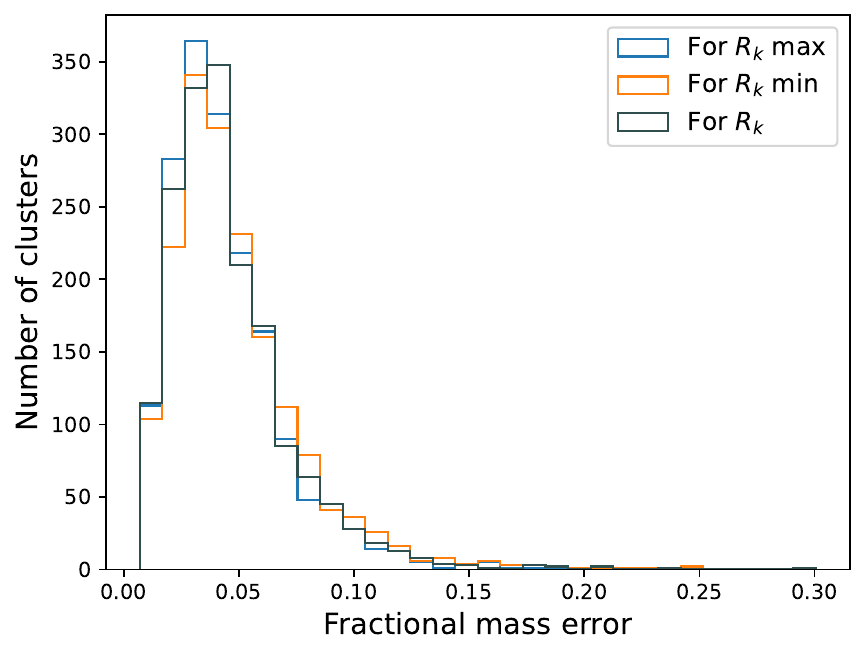}
    \caption{Distribution of the fractional mass error for masses determined using only stars inside $R_k$ (grey) and its lower and upper bound (orange and blue, respectively).}
    \label{fig:boot_mass_min_max}
\end{figure}

To estimate the fitting uncertainties of our mass determination method, we performed a bootstrap using 100 random samples of the magnitude distribution. We adopted the standard deviation as the error of our method for each cluster. Fig.~\ref{fig:boot_mass_min_max} displays the results for the fractional mass error, in the $G$ band, obtained using stars within $R_k$ and its lower and upper bounds. The majority of the clusters have a mass error below 4$\%$, with a standard deviation of 3$\%$. The mass errors within the lower and upper bound of $R_k$ are 3$\%$ and 5$\%$, respectively.

These results suggest that the uncertainty in the King radius does not significantly impact our mass determinations. It is important to note that these are only the statistical uncertainties associated with our method, not the true uncertainty of each cluster's mass. Systematic uncertainties related to age, distance and reddening determinations are not accounted for.

Regarding systematic differences arising from the choice of the IMF, we compared the masses obtained above with those derived using the \citet{Chabrier_2003} log-normal IMF. Within the magnitude limit and distance range of our sample, the mass differences range from approximately 10\% for closer clusters to 4\% for clusters beyond $\sim$1 kpc. This distance-dependent difference arises because lower-mass stars (the regime where the differences between the Chabrier and Kroupa IMFs are most pronounced) are more detectable at closer distances. The differences in OC masses obtained using the two IMFs provide an estimate of the systematic uncertainties involved. Further details on this comparison are provided in Appendix~\ref{ch:appendixC}.

Additionally, if we account for the escape of a fraction of low-mass stars due to mass-segregated evaporation, the systematic error could increase to approximately 20\%, assuming about half the stars below $\sim$0.4 $M_{\odot}$ have escaped. The exact value would depend on the cluster's dynamical state.

Finally, we also note that another source of uncertainty in our mass determinations arises from the contribution of binary stars. Binary fractions and mass ratios vary within the tidal radius, depending on the cluster's dynamical state \citep[e.g.,][]{2024MNRAS.528.6211A}, potentially further influencing the mass estimates. We proceed with these caveats in mind and provide further discussion on their impact on the OC dissolution timescale in Sect.~\ref{ch:conclusion}.

\subsubsection{Consistency assessment}
\begin{figure}[ht]
    \centering
    \includegraphics[width=0.95\linewidth]{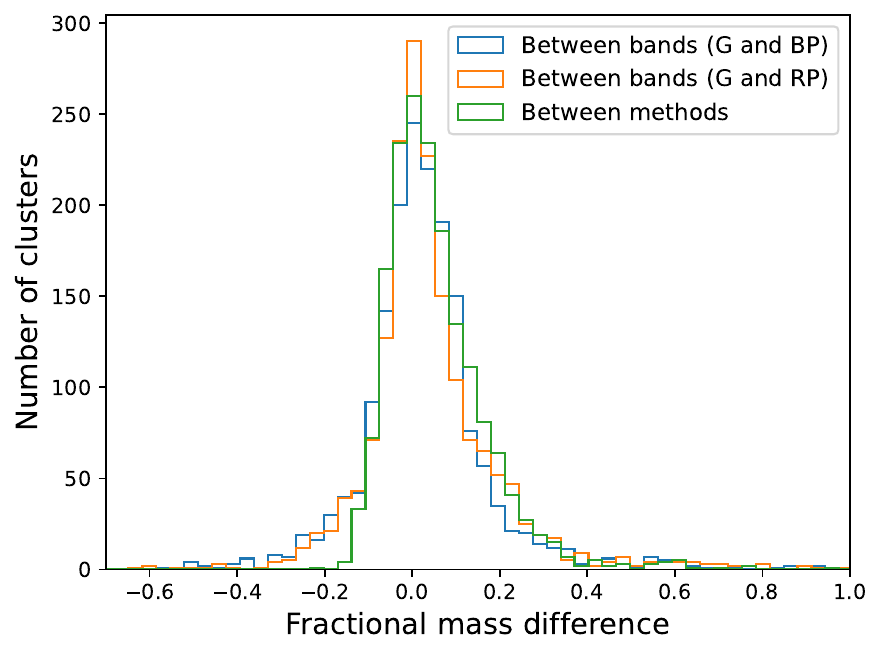}\  
    \caption{Fractional mass difference for the mass determined using the $G$ and $G_{BP}$ bands (blue), $G$ and $G_{RP}$ bands (orange) and using the adopted method and a sanity check, in green.}
    \label{fig:robust_mass}
\end{figure}

To further assess the consistency of our mass determination method, we determined the mass using the $G_{BP}$ and $G_{RP}$ \textit{Gaia} bands. Relative to masses determined from the $G$ band, there is a normalized median difference of 3$\%$ and 5$\%$ for $G_{BP}$ (Fig.~\ref{fig:robust_mass}, in blue) and $G_{RP}$ bands (Fig.~\ref{fig:robust_mass}, in orange), with associated standard deviations of 18$\%$ and 20$\%$, respectively. Essentially, the mass determinations using different bands yields similar results.

Additionally, we compared our primary mass determination method with the sanity check method described in Sect. \ref{ch:mass_det}, for the same band. The mean mass difference between the two methods (Fig. \ref{fig:robust_mass}, green) is 6$\%$, with a standard deviation of 13$\%$. Notably, our main method consistently yields lower mass estimates for the clusters. This method is more sensitive to the LF morphology, particularly for cases where cluster parameters such as age or distance are imprecisely determined or when there is data incompleteness. In contrast, the validation method is less susceptible to these factors as previously mentioned.

The observed mass discrepancies between methodologies and bands align closely with the typical mass error of 6\%, affirming the robustness of our mass determination approach. Given that the results obtained from $G_{BP}$ and $G_{RP}$ bands were only used as a validation check, $G$ band-derived masses are regarded as the main estimations.
 
\subsubsection{Mass classification} 

As a further validation step, in order to evaluate the reliability of mass determinations and exclude low-quality results, we classified the 1724 open clusters based on the quality of the agreement between observed and model luminosity distributions.  A preliminary sorting of clusters was performed  using the root mean square (RMS) of the fit, followed by a visual classification into three categories: M1 (best), M2 and M3 (worst), as shown in Table \ref{tab:M_class}. 

\begin{figure}[ht]
    \centering
    \includegraphics[width=0.9\linewidth]{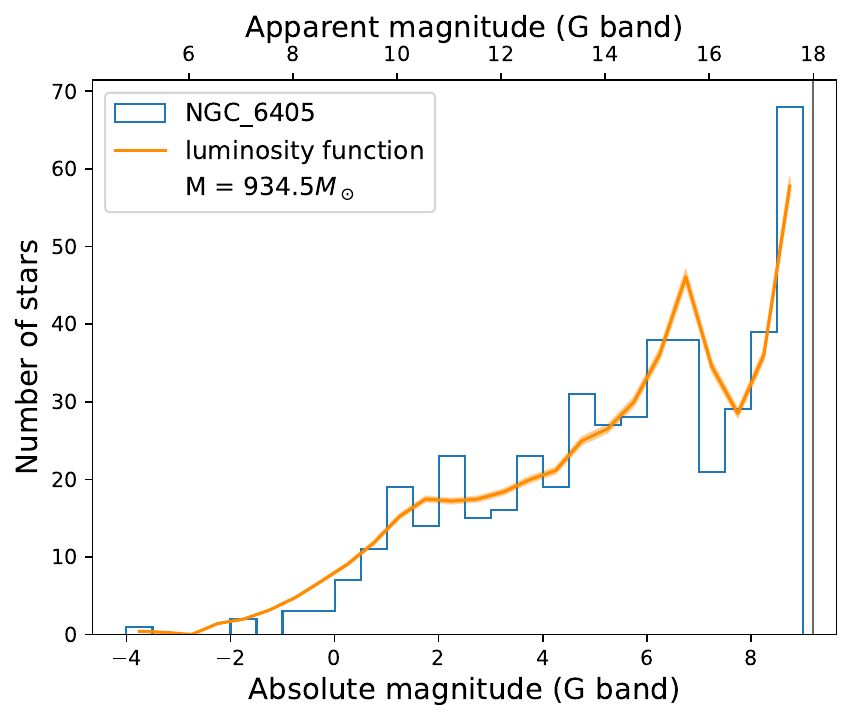}\  
    \caption{Magnitude distribution of NGC~6405 (for stars with membership > 0.7) with luminosity function (orange line) scaled to match the observed density distribution (blue histogram). The error is displayed as the shaded orange area. The bin width is 0.5 pc.}
    \label{fig:M1_ex}
\end{figure}

Clusters assigned to the M3 category have mass determinations deemed unreliable, with significant disparities between the observed magnitude distribution and the theoretical LF. Additionally, a supplementary category, denoted as MX, was established for clusters featuring poorly populated or undefined colour-magnitude diagrams. This classification includes all OCs previously categorized as P3. Clusters assigned to the MX category have poor quality CMDs, rendering their mass determinations invalid. An illustrative example of an M1 classification is presented in Fig. \ref{fig:M1_ex}.

\begin{table}[ht]
\caption{Number of clusters per classification according to the quality of the mass determination.}
\centering
\begin{tabular}{lll}
\hline\hline
Class & Number & Fraction  \\ \hline
M1 & 812 & 47.1 $\%$\\ 
M2 & 695 & 40.3 $\%$ \\ 
M3 & 65 & 3.8 $\%$ \\ 
MX & 152 & 8.8 $\%$  \\ \hline
\end{tabular}
\label{tab:M_class}
\end{table}

\subsubsection{Comparison with other studies \label{ch:compare_mass}}

\begin{figure}[ht]
    \centering   
    \includegraphics[width=0.95\linewidth]{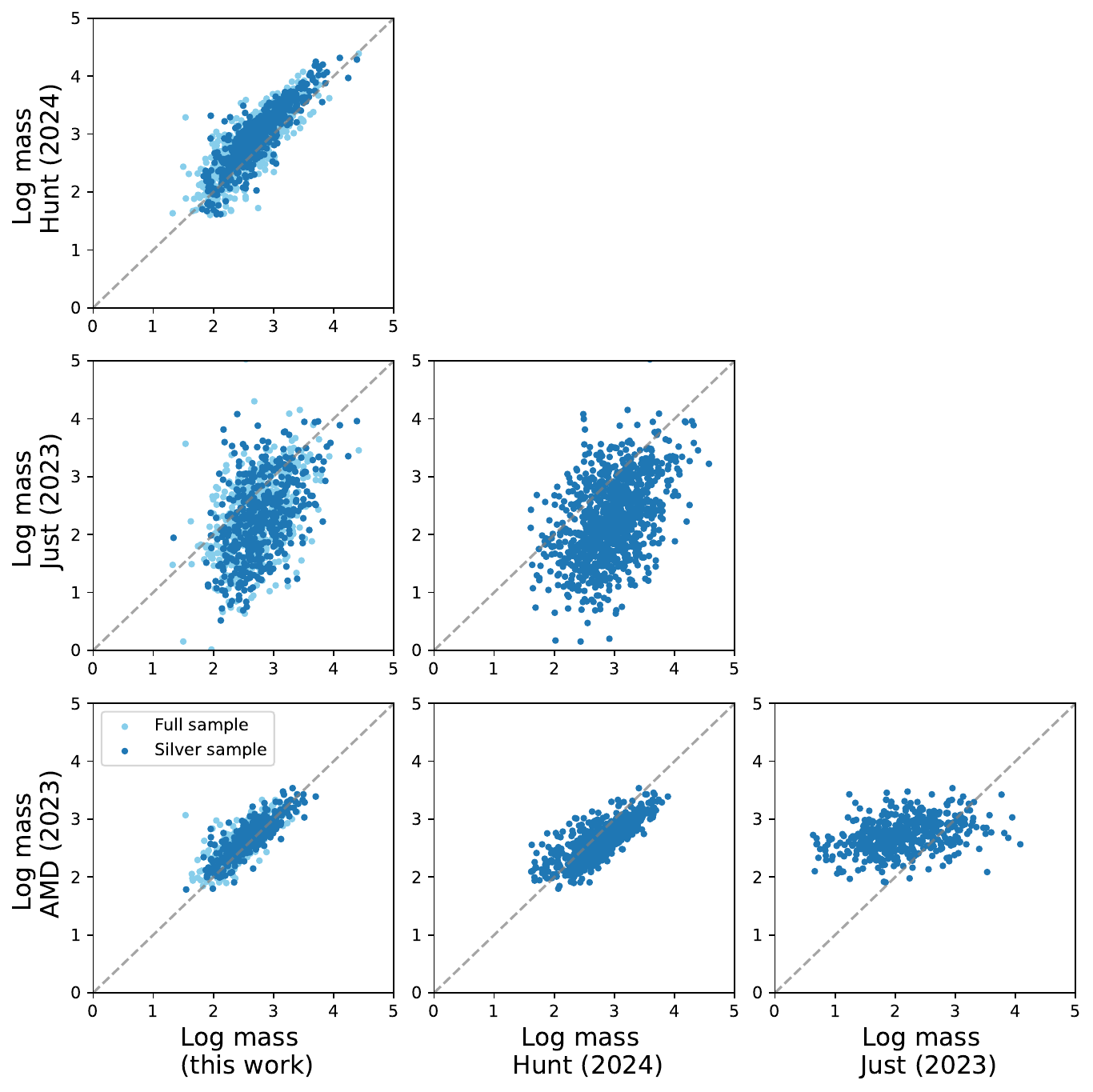}
    \caption{Comparison of individual masses from  \citetalias{anderson_2023}, \citet{Just_2023} and \citet{hunt_III_masses}  with the masses determined in this work.}
    \label{fig:compare_mass}
\end{figure}

Fig.~\ref{fig:compare_mass} presents a comparison of our results with the available large catalogues of OC masses from \citet{Just_2023}, \citetalias{anderson_2023}  and \citet{hunt_III_masses}. In the last two catalogues, masses were derived by integrating the mass function constructed using individual stellar masses. The first catalogue provides tidal masses, estimated from the galactic tidal forces exerted on the cluster limiting radii, as detailed in \citet{piskunov_tidal_2008}. 

There is a general agreement with the photometric masses from \citetalias{anderson_2023} and \citet{hunt_III_masses}, despite being determined using a different method. We observe though a tendency for higher masses in \citet{hunt_III_masses}, especially in the massive end. This tendency had been noted by \citet{hunt_III_masses} in their comparison with the masses from \citetalias{anderson_2023} and which they ascribed to the additional corrections for Gaia
incompleteness by \citet{hunt_III_masses}. Given that \textit{Gaia} DR2 is essentially complete for the $G < 18$ mag cut adopted here and by \citetalias{anderson_2023}, it is possible that \citet{hunt_III_masses} have slightly over-corrected their OC masses, which employ fainter stars.

However, comparing with \citet{Just_2023}, individual masses display notable discrepancies for most clusters. We note that their (tidal) masses are proportional to $r_{\text{tidal}}^3$, which makes them highly affected by uncertainties in the tidal radius, Specifically, their error distribution exhibits a mean value of 93\% with a standard deviation of 48\% \citep[see Fig. 5 of ][]{Just_2023}.
For these reasons, and at least until better radius determinations become available, we consider the luminous masses more reliable.

\begin{figure}[ht]
    \centering  
    \includegraphics[width=0.95\linewidth]{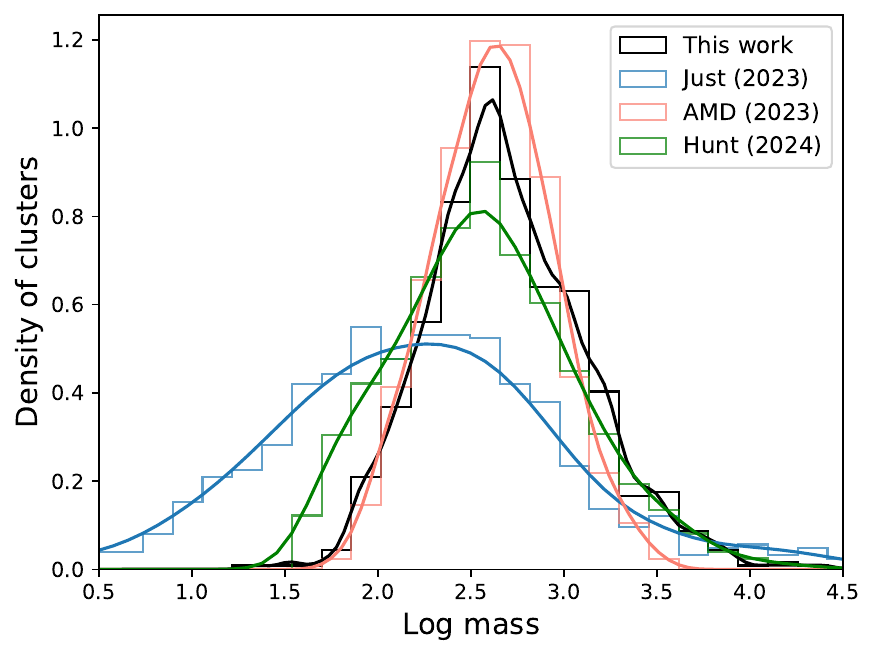}
    \caption{Mass distributions from this work , \citetalias{anderson_2023}, \citet{Just_2023} and \citet{hunt_III_masses}, with KDEs represented as as solid lines. }
    \label{fig:compare_mass_hist}
\end{figure}

Fig.~\ref{fig:compare_mass_hist} shows the distribution of masses of the four catalogues. Again, a general agreement with \citetalias{anderson_2023} and \citet{hunt_III_masses} is observed, with the mass distributions peaking at $\sim$ 2.6-2.7. We also note that with respect to this study, the mass distribution of \citetalias{anderson_2023} is tighter and that of \citet{hunt_III_masses} is broader. The tidal masses of \citet{Just_2023} follow a much broader distribution with the peak at lower masses.

\subsection{Sample selection \label{ch:sample_select}}

Given the quality classifications described above, we established two sub-samples. The first, labelled the "gold sample", contains only clusters classified as best quality across all three categories (P1, R1, M1). The second sample, denoted as the "silver sample", includes clusters with intermediate to high-quality classifications, i.e., classifications 1 and 2 for photometry, radius and mass. The gold sample contains 153 OCs (9$\%$), while the silver sample contains 713 OCs (41$\%$). It should be emphasized that the gold sample is entirely contained within the silver sample. 

\begin{figure}[ht]
    \centering
    \includegraphics[width=0.9\linewidth]{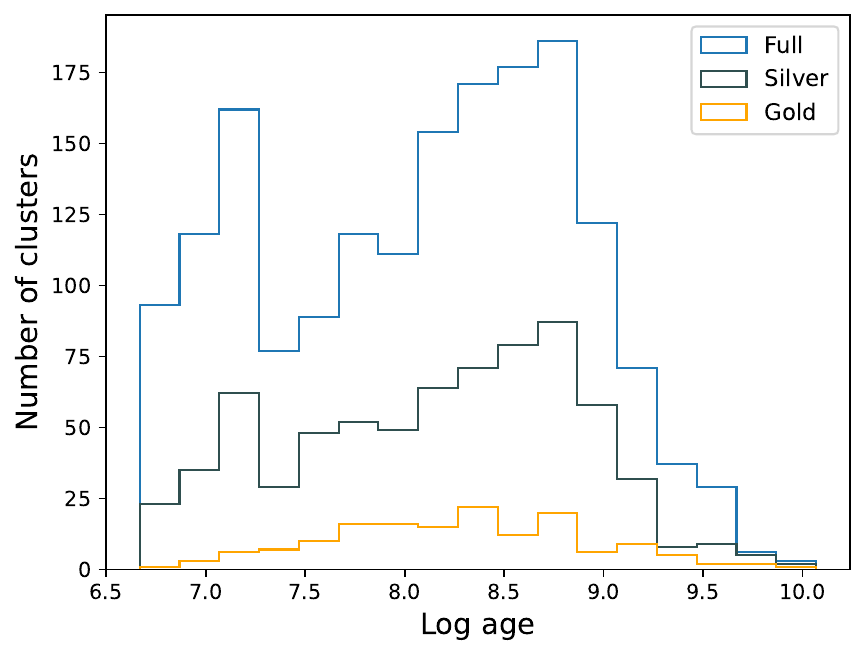}\
    \includegraphics[width=0.9\linewidth]{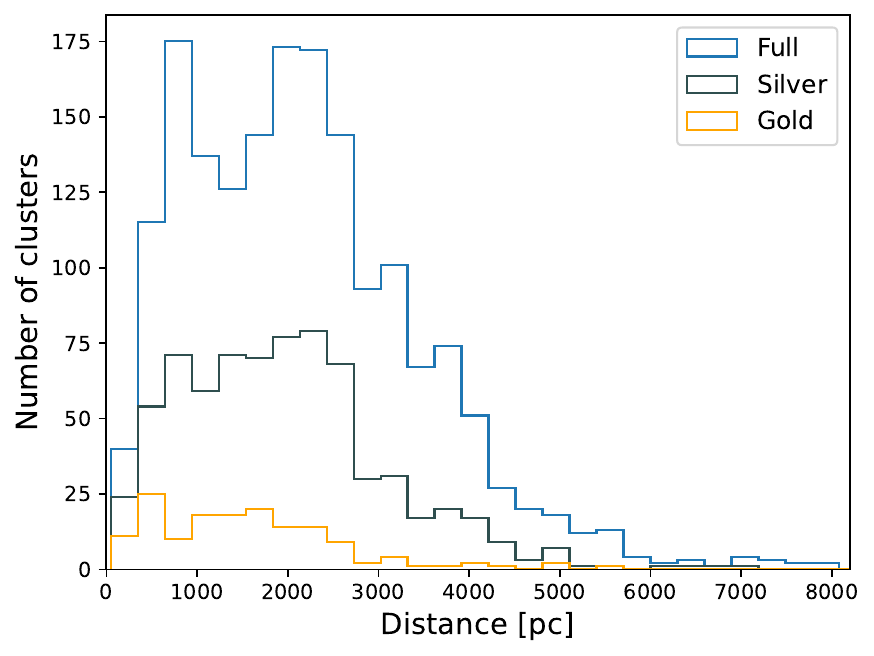}
    \caption{Distribution of age (top) and distance (bottom) for the full sample (blue), silver sample (grey) and gold sample (orange).}
    \label{fig:samples_dist_age}
\end{figure}

The age and distance distributions of each sample are shown in Fig.~\ref{fig:samples_dist_age}. These distributions indicate that the silver sample is distributed similarly to the full sample, indicating the absence of apparent biases resulting from the quality-based selections used to generate the silver sample. In contrast, the gold sample is manifestly small, displaying a much more limited spatial coverage. For these reasons, we adopt the silver sample as our baseline for investigating the OC mass distribution and disruption timescale in the solar neighbourhood.

The age distribution in Fig.~\ref{fig:samples_dist_age} shows a marked decline in the number of OCs older than approximately 1 Gyr. While this decrease is expected due to cluster disruption, we cannot rule out that part of it is due to the incompleteness of older, fainter clusters \citep{Moitinho_2010, sandro_2024}. Therefore, we limit our analysis to OCs up to 1 Gyr. 

Another prominent feature in the age distribution is the peak at young ages, around log(age) $\sim$ 7.1. This peak is also seen in the catalogue of \citet{cantat-gaudin_painting_2020}, but it is not observed in the catalogues of \citet{Hunt_II_2023} and \citet{lorenzo_quad_2024}. The reality of this peak is unclear, with \citet{anders_star_2021} attributing it to an increased cluster formation rate 6-20 Myr ago, while \citet{lorenzo_quad_2024} attribute it to an artifact caused by red and faint contaminants in isochrone fits. Additionally, $\sim$ 20 Myr corresponds to the end of violent relaxation and the rapid dispersal associated with the gas expulsion phase of young clusters \citep{Shukirgaliyev_2019}, which could lead to the observed decrease in the number of clusters just after 20 Myr. To avoid this source of uncertainty, we will consider only clusters older than 20 Myr when analysing cluster dispersion.

\begin{figure}[ht]
    \centering
    \includegraphics[width=0.95\linewidth]{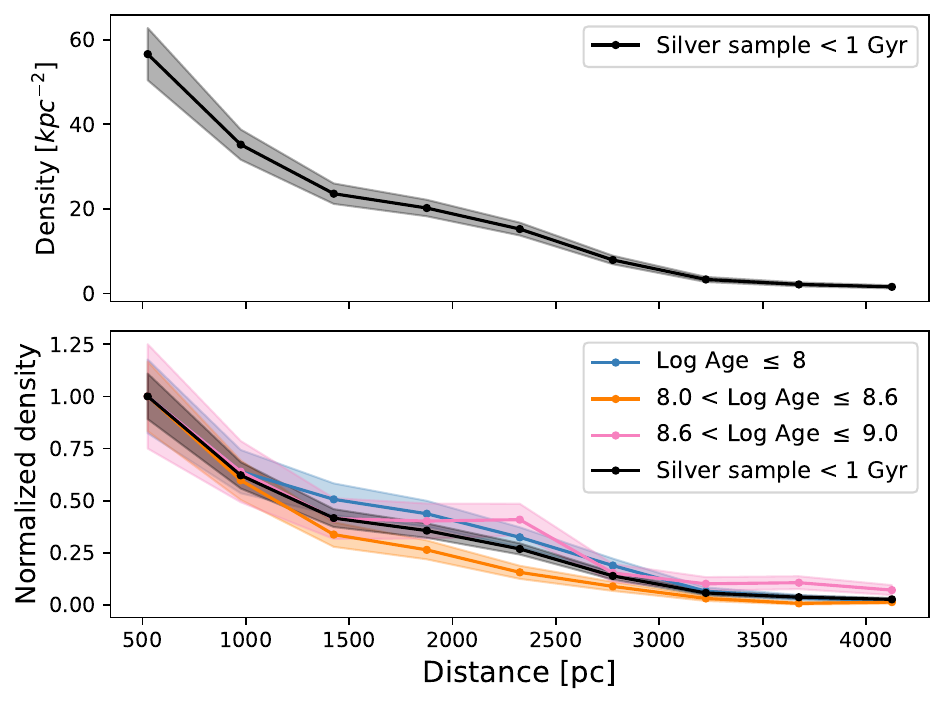}
    \caption{Radial density of the OCs in the silver sample with age under 1 Gyr (top panel) and for each age subsample, normalized to 1 at the maximum (bottom panel). Poisson errors are represented as shaded areas.}
    \label{fig:comp_1G}
\end{figure}

\begin{figure}[ht]
    \centering
    \includegraphics[width=0.95\linewidth]{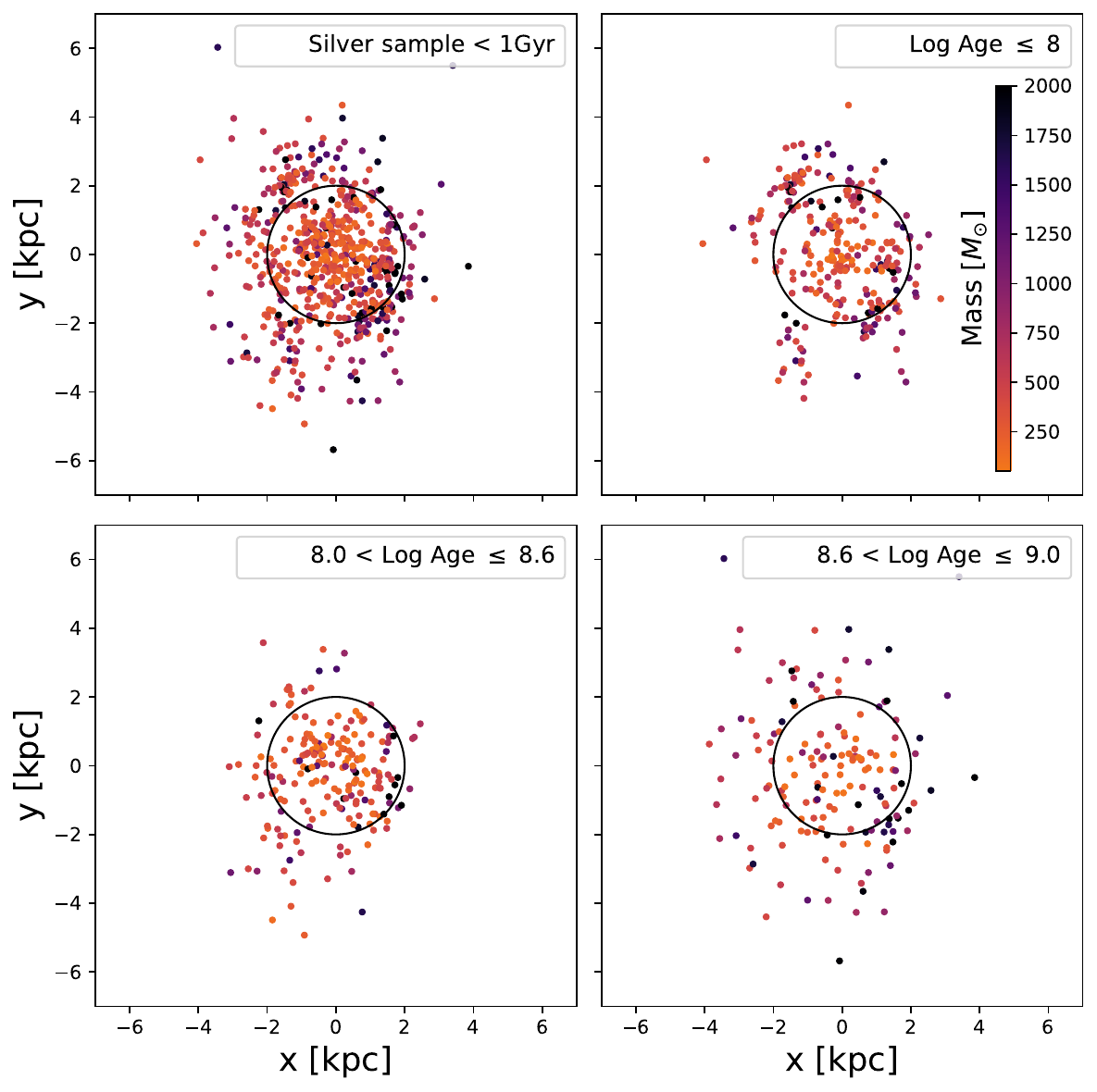}
    \caption{Spatial distribution of the silver sample (top left) and the 3 age subsamples with black circle at 2 kpc, projected on the Galactic plane. The OCs are colour-coded by mass. The X axis points to the galactic centre and Y points in the direction of rotation of the Galaxy. The Sun is located at (0,0).}
    \label{fig:comp_age_samp}
\end{figure}

We now analyse the spatial completeness of the silver sample following the approach in \citet{sandro_2024}.  For this, we divide the sample in concentric rings with steps of 450 pc and in three age groups:
log(age) $\leq$ 8.0, 8.0 $<$ log(age) $\leq$ 8.6 and 8.6 $<$ log(age) $\leq$ 9.0. The age limits were chosen to have a balanced number of clusters in each group. 
The radial density profiles for each group is displayed in the bottom panel of Fig. \ref{fig:comp_1G}. The distributions on the Galactic plane are presented in Fig. \ref{fig:comp_age_samp}.

In a complete sample, we would expect the density to remain roughly constant in our immediate neighbourhood. However, as shown in the top panel of Fig. \ref{fig:comp_1G}, the density decreases with increasing distance. 
The trend is also observed for the full sample, as thoroughly discussed in \citet{sandro_2024} and persists in our silver sample. 

At young ages, OCs are well known tracers of spiral arms \citep[e.g.][]{dias_lepine_2005, cantat-gaudin_painting_2020}, and thus present a structured distribution, as seen in Fig.~\ref{fig:comp_age_samp}. At intermediate and higher ages, the structure fades and the distribution appears more homogeneous.
The clumpy structures and irregular density profiles for younger clusters make it hard to assess completeness. Nevertheless, given the large variations seen in the density profiles (Fig.~\ref{fig:comp_1G}), observing consistent decreases in density profiles across all age subsets suggests uniform selection effects across different ages. 
From the spatial distributions (Fig.~\ref{fig:comp_age_samp}), we observe that beyond a 2 kpc limit, the distributions include regions that are no longer sampled, with larger radii featuring empty areas. For this reason, we restrict our analyses to within 2 kpc.

It is important to emphasize that, since the density profiles of the different age groups exhibit a similar decrease with the (cylindrical) distance to the Sun, their ratios remain largely unaffected by incompleteness. We note that we are not referring to the absolute numbers of clusters, but to the rate at which they decrease with distance. 
Although the samples are incomplete, the fraction of clusters removed by incompleteness appears to be independent of age. This is a relevant point because the effects we are studying, such as the disruption rate, are expressed as ratios of cluster numbers across different age groups. 
Consequently, this particular source of incompleteness should not influence the determination of the dissolution rate. Nonetheless, the distance cut-off is necessary to mitigate sampling errors in the low-count regime.

\begin{figure}[ht]
    \centering                  
    \includegraphics[width=0.95\linewidth]{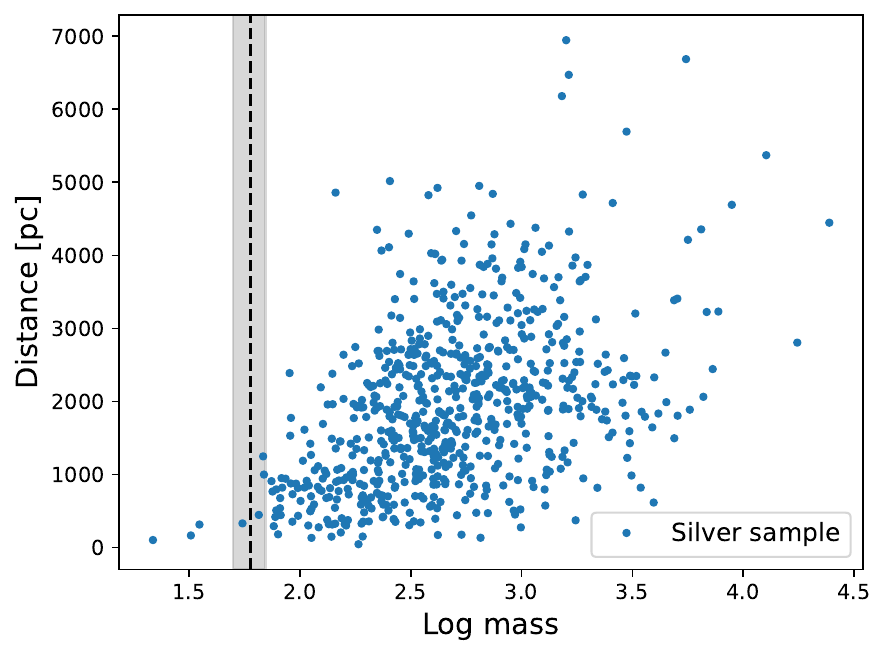} 
    \caption{Distribution of distances as a function of the logarithm of mass, for the 713 OCs in the silver sample. The grey shaded area represents the interval of masses from 50 to 70 $M_\odot$ with the black dotted line at 60 $M_\odot$.} 
    \label{fig:mass_vs_dist}
\end{figure}

We now inspect the mass distribution of the silver sample at different distances presented in Fig.~\ref{fig:mass_vs_dist}. A barrier at 60 $M_\odot$, separating the bulk of the clusters from 3-4 low mass and sparsely distributed very close clusters is readily seen. This suggests that a $\sim 60 M_\odot$ lower mass limit below which stars cannot remain bound in an OC in the solar neighbourhood. The 3-4 low mass objects have few stars, which exhibit a super virial velocity dispersion, thus likely being poor OCs in the final phases of dissolution only seen due to their closeness. These objects are, from nearest to farthest, Alessi\_13, UPK\_606, UPK\_385, and UPK\_624.  A similar mass barrier is also seen in the data of \citetalias{anderson_2023}, who do not include the 3 low mass outliers. We note that a selection effect would likely result in a smoother distribution, approximately triangular, extending toward smaller masses across the barrier. However, at this stage, we view the $\sim 60 M_\odot$ minimum mass more as a well-motivated indication than a definitive claim. A robust determination will require further investigation.

In summary, the curated sample that will be used below in our model is the silver sample (all quality flags = 1 or 2), restricted to clusters within 2 kpc of the Sun, with ages between 20 My and 1 Gyr, and masses greater than 60 $M_\odot$. Tab.~\ref{tab:OC_results}\footnote{The full table with all 1724 OCs will be available at the CDS.} lists the studied OCs, their determined radii, masses, uncertainties, and quality flags.

\begin{table*}[t]
    \caption{\label{tab:OC_results} Determined radii, masses, uncertainties, and quality flags for 10 randomly selected OCs.}
    \centering
    \begin{tabular}{lrrrrrrrrrrr}
        \hline\hline
        Name & $r_c$ (pc) & $\sigma^{-}_{r_c}$ (pc) & $\sigma^{+}_{r_c}$ (pc) & $r_k$ (pc) & $\sigma^{-}_{r_k}$ (pc) & $\sigma^{+}_{r_k}$ (pc) & $M(r_k)$ $(M\textsubscript{$\odot$})$ & $\sigma_{M}$ $(M\textsubscript{$\odot$})$  & $M_{f}$ & $P_{f}$ & $R_{f}$ \\
        \hline
        UPK 640 & 3.58 & 0.32 & 0.47 & 12.38 & 2.51 & 3.27 & 419.32 & 8.24 & 1 & 1 & 1 \\
        Alessi 12 & 8.67 & 1.78 & 1.26 & 11.32 & 0.86 & 2.18 & 416.12 & 12.14 & 1 & 1 & 2 \\
        Roslund 3 & 1.01 & 0.09 & 0.12 & 37.49 & 25.66 & 40.20 & 399.38 & 13.25 & 1 & 1 & 2 \\
        NGC 3572 & 3.92 & 1.80 & 1.55 & 7.33 & 1.45 & 30.88 & 659.22 & 30.74 & 2 & 2 & 2 \\
        King 19 & 1.80 & 0.40 & 0.98 & 7.00 & 3.21 & 26.37 & 1100.31 & 31.83 & 1 & 2 & 1 \\
        NGC 6268 & 1.38 & 0.16 & 0.18 & 5.98 & 1.55 & 3.24 & 447.01 & 15.93 & 2 & 1 & 1 \\
        NGC 6259 & 2.24 & 0.11 & 0.12 & 31.08 & 9.71 & 21.27 & 5055.92 & 51.55 & 2 & 2 & 1 \\ 
        IC 1369 & 1.26 & 0.11 & 0.14 & 6.71 & 1.59 & 2.91 & 1374.01 & 35.58 & 1 & 2 & 1 \\
        Haffner 8 & 3.51 & 0.99 & 0.72 & 4.83 & 0.69 & 1.82 & 553.01 & 15.84 & 1 & 1 & 2 \\
        Berkeley 39 & 4.51 & 0.47 & 0.70 & 11.47 & 1.71 & 2.21 & 6456.33 & 167.55 & 1 & 1 & 1 \\
        \hline
    \end{tabular}

    \tablefoot{Columns 2-7 include the structural parameters: core radius ($r_c$) and King radius ($r_k$) with their respective lower and upper uncertainties ($\sigma^{-}$, $\sigma^{+}$, respectively). The mass within the King radius ($M(r_k)$) with its uncertainty ($\sigma_{M}$) are in columns 8-9. The visually assigned classifications for the mass, photometry and radii ($M_{f}$, $P_{f}$, $R_{f}$, respectively) are in columns 10-12. We also include the positions, ages, distances and $A_v$ from the \citet{dias_updated_2021} catalogue, and the values with respective uncertainties for parameters $c$ and $N_0$ in the full table. An additional column with the assigned sample is also provided (2 = gold sample; $\geq$ 1 silver sample; $\geq$ 0 full sample). The full table with the 1724 OCs is available at the CDS.} 
\end{table*}

\section{Model for the population of dissolving OCs \label{ch:disruption}}

Previous studies have shown that the disruption time, $t_{dis}$, of an OC of mass $M_i$, can be described by $t_{dis} = t_0(M_i/M_\odot)^{\gamma}$ with a timescale $t_0$ \citep{boutloukos_star_2003}. This timescale depends on the tidal forces of the host galaxy, so it varies between galaxies and is dependent on the position of the cluster in the galaxy. The disruption time ($t_{dis}$) increases with the cluster initial mass \citep{lamers_analytical_2005} and this dependence is quantified by the parameter $\gamma$, which depends on the concentration of the stellar distribution in the cluster \citep{lamers_mass_2010}.

The strength of the tidal forces increases with proximity to the Galactic centre, making the intensity of disruption dependent on the location of the cluster within the galaxy. However, as the clusters in our sample are located within a few kiloparsecs around the Sun, we will assume the same tidal influence for all clusters. The parameter $\gamma$ is also assumed to be the same for all clusters, as we do not consider the differences in stellar concentration in the clusters. 
While estimates for parameters $\gamma$ and $t_0$ have been made in several studies analysing the OC age distribution, \citep[e.g.][]{boutloukos_star_2003, baumgardt_dynamical_2003, lamers_analytical_2005}, to the best of our knowledge, the mass distribution has not been considered in previous studies.

To simulate the dissolution rate experienced by open clusters in the solar neighbourhood, we follow \citet{lamers_analytical_2005} employing a simple model that simulates the build up and mass evolution of a population of OCs along time. The model assumes a constant cluster formation rate and draws their initial masses from an Initial Cluster Mass Function (ICMF). As in \citet{lamers_analytical_2005}, we adopt adopting a power-law ICMF \citep{lada_embedded_2003}: $dN/dM \propto M^{-\alpha}$ with $\alpha \sim 2$, $M_{min}$ = 100 $M_{\odot}$ and $M_{max}$ = $3 \times 10^4$ $M_{\odot}$. Later in Sect.~\ref{ch:disruption_timescale} we will discuss the implications of this choice and explore the adoption of other ICMF functionals.

For the dissolution process, \citet{lamers_clusters_2006} provide 4 separate equations for the OC mass loss due to: stellar evolution, disruption by the galactic tidal field, spiral arm shocking and molecular cloud encounters. However, our work is focused on the total disruption of the OCs and we do not distinguish between different disruption mechanisms. For this reason, we combine the equations from \citet{lamers_clusters_2006} in a single equation that reproduces the overall mass dependent disruption time.

The expression for the total mass loss is: 
\begin{equation}
    \frac{dM}{dt} = - \frac{\left(M_i\right)^{1-\gamma} (10^4)^\gamma }{t_4^{\text{tot}} \gamma}
    \label{eq:miu}
\end{equation}
where $M_i$ is the cluster initial mass in the simulation. This equation reproduces the same mass loss over time as the 4 separate equations from \citet{lamers_clusters_2006} with $t_4^{\text{tot}}$ = 1.87 Gyr and $\gamma$ = 0.67. As described in \citet{lamers_analytical_2005}, $t_0 \sim \gamma \ t_4^{\text{tot}}$ so the disruption timescale can also be defined as $t_{dis} = t_4^{\text{tot}} \ \gamma \ (M_i/ 10^4)^{\gamma}$ which results in Eq. \ref{eq:miu}. 

\section{Comparison with observations \label{ch:disruption_timescale}}

We start by running simulations in coarse 10x9 grids, with $t_4^{\text{tot}}$ ranging from 1 to 10 Gyr in steps of 1 Gyr and $\gamma$ ranging from 0.1 to 0.9 in steps of 0.1. To quantify the agreement between the simulations and the observational sample (Sect.~\ref{ch:sample_select}) we calculated the likelihood between the predictions and observations, normalizing the number of simulated clusters to match the number of observed OCs. We consider the total likelihood as the sum of the log-likelihood from the age and mass distributions, separately. For the standard deviation, we consider Poisson errors $\sqrt{N}$, where N is the number of OCs in each bin.

\subsection{Results with the power-law ICMF}

\begin{figure}[ht]
    \centering
    \includegraphics[width=0.9\linewidth]{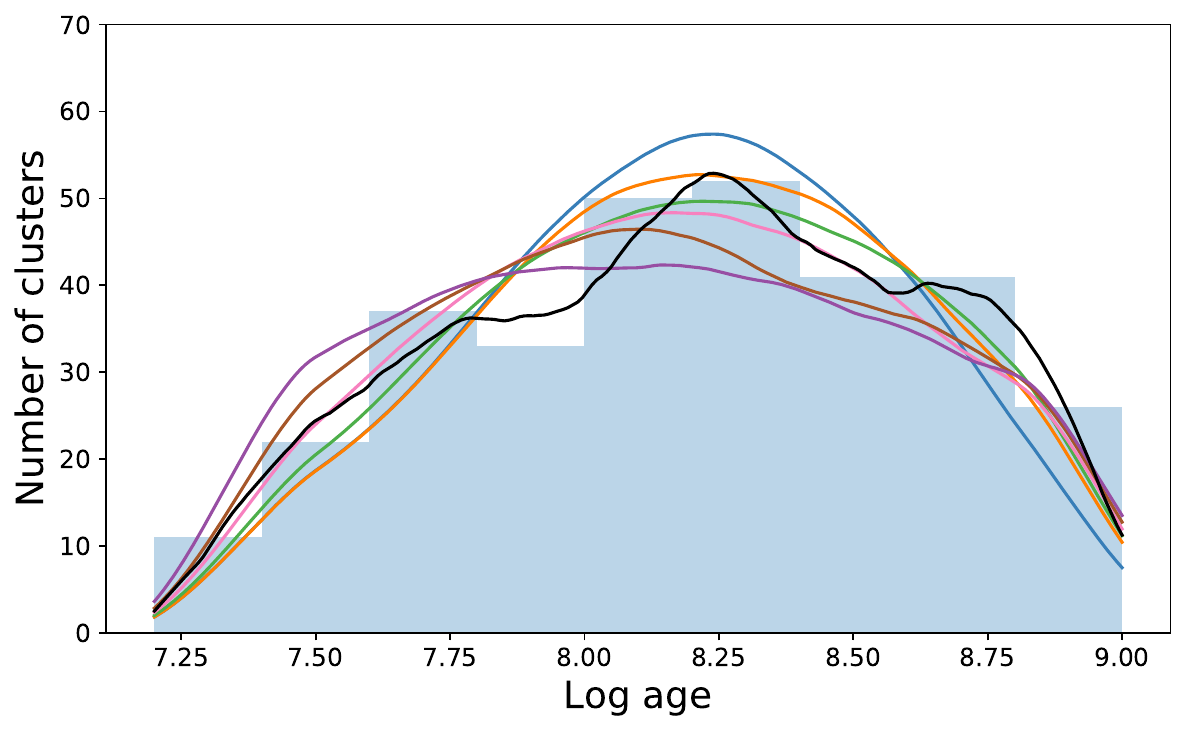}\
    \includegraphics[width=0.9\linewidth]{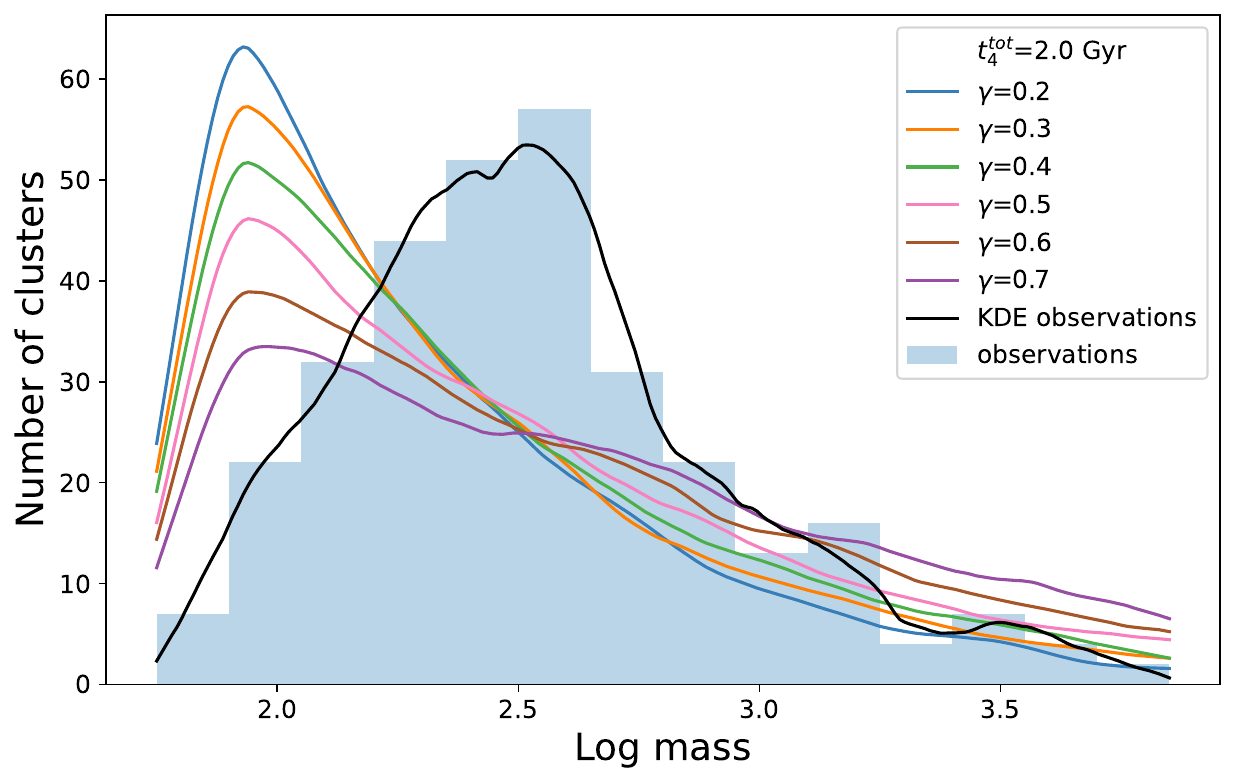}
    \caption{Comparison of the distribution of age (top) and mass (bottom) between the simulations (colour-coded KDEs) and the observations (blue histogram). For the simulations, we considered the power-law ICMF, $t_4^{\text{tot}}$ = 2 Gyr and $\gamma$ from 0.2 to 0.7. The observations are from the silver sample within 2 kpc around the Sun, with the mass and age cuts mentioned in the text. The solid black line is the KDE of the observations. All KDEs use the Epanechnikov kernel.}
    \label{fig:DIST_silver_2kpc_power}
\end{figure}

The optimal range of values for the parameters $t_4^{\text{tot}}$ and $\gamma$ considering the age distribution are obtained for a region where $t_4^{\text{tot}}$ = 2 Gyr with $\gamma$ between 0.2 and 0.7 and $t_4^{\text{tot}}$ = 3 Gyr with $\gamma$ between 0.5 and 0.8. We note that the $t_4^{\text{tot}}$ value we find is larger than the 1.3 Gyr-1.7 Gyr reported in the literature \citep{lamers_analytical_2005, lamers_clusters_2006, anders_star_2021}. In the top panel of Fig.~\ref{fig:DIST_silver_2kpc_power}, we present the observed and simulated age distributions for different values of $\gamma$ given a fixed $t_4^{\text{tot}}$ of 2 Gyr where we observe a good agreement.

In contrast, for the masses it is not possible to isolate an optimal region in the grid. In fact, although the region defined above provides good results for the age distribution, the simulated mass distribution does not match the observations, for any of the combinations considered. This effect is illustrated in the bottom panel of Fig. \ref{fig:DIST_silver_2kpc_power} where the observed mass distribution is shown with the KDEs of the simulated distributions for different values of $\gamma$ given a fixed $t_4^{\text{tot}}$ of 2 Gyr. It is clear that, independently of the value of $\gamma$, the simulations do not reproduce the observed mass distribution despite the general good agreement with the ages. For a fixed $\gamma$, the effect of changing $t_4^{\text{tot}}$ in the mass distribution is small so none of the combinations in the grid reproduces the observed masses.

This incompatibility was unexpected and was not suggested in previous works \citep[e.g.][]{lamers_clusters_2006, lamers_disruption_2005}. It has been revealed by our new mass catalogue and clearly indicates the need for adjustments in the physical ingredients of the model. We also checked the results using the silver sample within 1.5 kpc and the gold sample within 2 kpc, arriving at the same conclusion.

The two key factors influencing the overall shapes of the age and mass distributions are the cluster formation rate and the ICMF. Several studies \citep[e.g.][]{Snaith_2015_sfr}  indicate that the solar neighbourhood experienced a relatively uneventful star formation history over the past Gyr. Given this, along with the good agreement observed in the age distribution, we find that the discrepancy likely lies in the ICMF.

\subsection{Exploring alternative forms for the ICMF}

The power-law ICMF of \citet{lada_embedded_2003} is widely adopted for simulating the initial mass distribution of OC populations in the Milky Way. In particular, it was adopted in \citet{lamers_clusters_2006, lamers_disruption_2005} for studying the disruption timescales of OC. However, this ICMF was determined for embedded clusters, of which a large fraction will not survive the gas expulsion phase \citep[e.g.][]{lada_embedded_2003, BaumgardtKroupa2007, Shukirgaliyev_2017}. In this view, it cannot  be taken for granted that the surviving open clusters will have the same mass distribution when they emerge from their parent molecular clouds. 

Indeed, \citet{parmentier_2008_icmf} address the difference between the embedded and non embedded cluster mass functions. They define the ICMF as the mass function of star clusters right after the effects of gas expulsion have ended \citep[$< 20$ Myr][]{Shukirgaliyev_2017, Shukirgaliyev_2019}. In their model, they find that the ICMF can differ from the embedded counterpart, with the shape depending on the star formation efficiency. For a (mass-independent) star formation efficiency of 20\% the embedded power-law mass function becomes a bell-shaped ICMF. At efficiencies $\sim$ 40\% the shape of the embedded mass function is preserved.

In the study of the minimum mass of bound clusters in different galaxies,  \citet{trujillo_2019} describe the (non-embedded) ICMF as a power-law ICMF with an exponential truncation at both high- and low-mass ends, effectively creating a bell like shape: 
\[
\frac{dN}{dM} \propto M^\beta \exp\left(-\frac{M_{\text{min}}}{M}\right) \exp\left(-\frac{M}{M_{\text{max}}}\right)
\]
with  $\beta$ = -2. For the solar neighbourhood they find that $M_{\text{min}} = 1.1 \times 10^{2} \, M_{\odot}$ and $M_{\text{max}} = 2.8 \times 10^{4} \, M_{\odot}$ which are the minimum and maximum mass scales, respectively.

\begin{figure}[ht]
    \centering
    \includegraphics[width=0.9\linewidth]{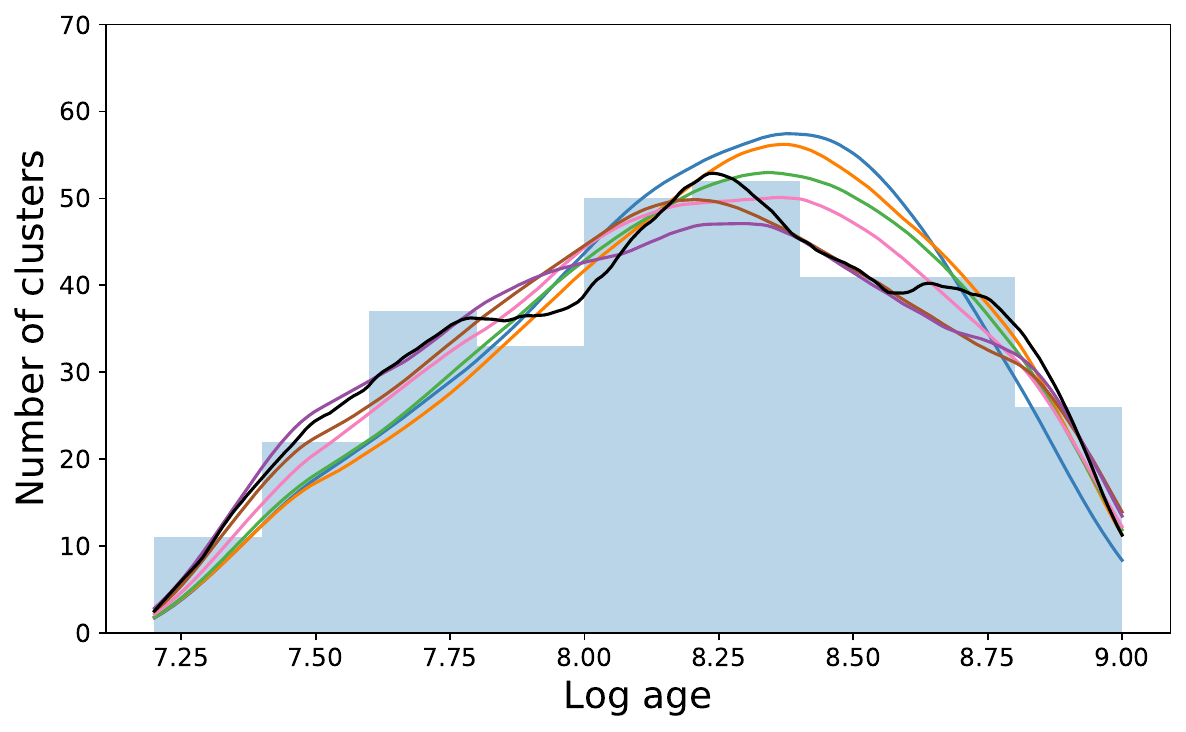}\
    \includegraphics[width=0.9\linewidth]{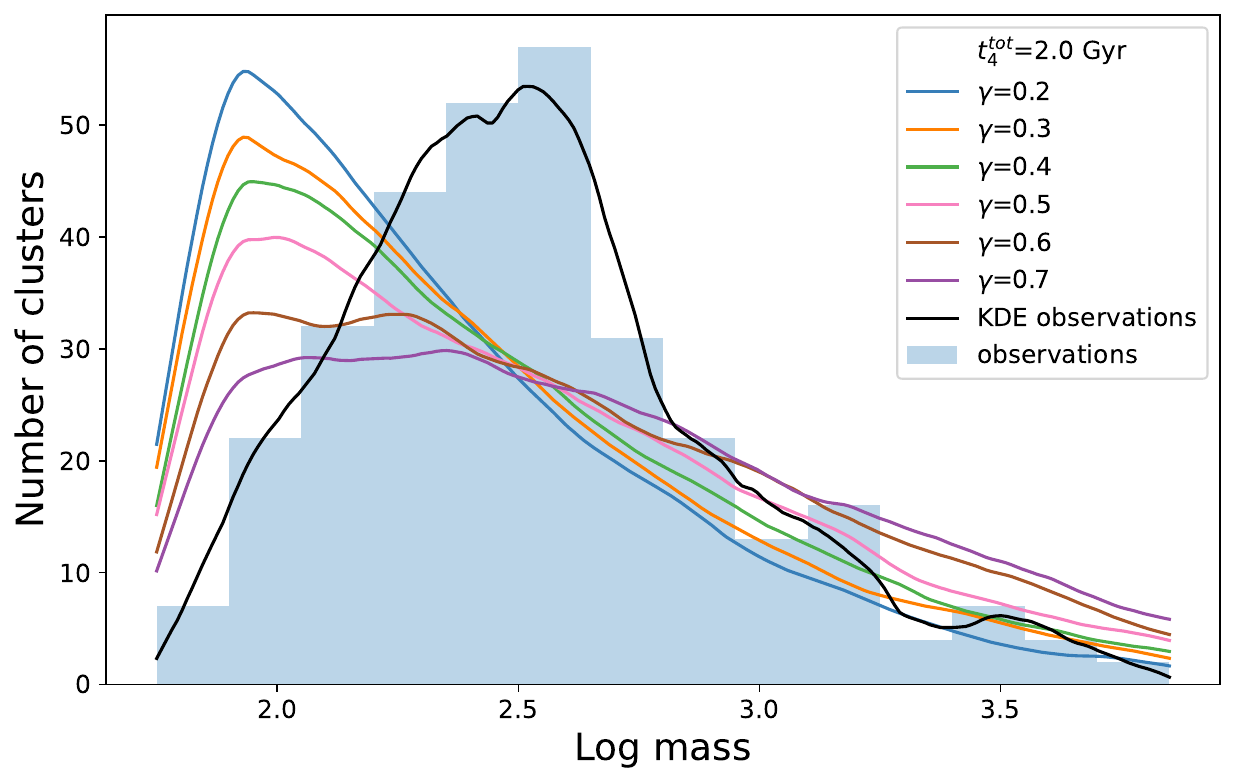}
    \caption{Comparison of the distribution of age (top) and mass (bottom) between the simulations (colour-coded KDEs) and the observations (blue histogram with fitted black KDE) considering $t_4^{\text{tot}}$ = 2 Gyr and $\gamma$ from 0.2 to 0.7, using the ICMF from \citet{trujillo_2019}.}
    \label{fig:DIST_silver_2kpc_trujillo}
\end{figure}

The effect of adopting the ICMF proposed by \citet{trujillo_2019} in our model is presented in Fig.~\ref{fig:DIST_silver_2kpc_trujillo}. While we find good fits to the age distribution, we observe that this ICMF also does not reproduce the observed mass distribution. We note, however, that  \citet{trujillo_2019} set their ICMF parameters  $\beta$,  $M_{\text{min}}$, and $M_{\text{max}}$ using pre-\textit{Gaia} masses from \citet{lamers_analytical_2005} based on the \citet{Kharchenko_2005} OC catalogue (plus unpublished masses provided by H. Lamers, private communication).  

\begin{figure}[ht]
    \centering
    \includegraphics[width=0.95\linewidth]{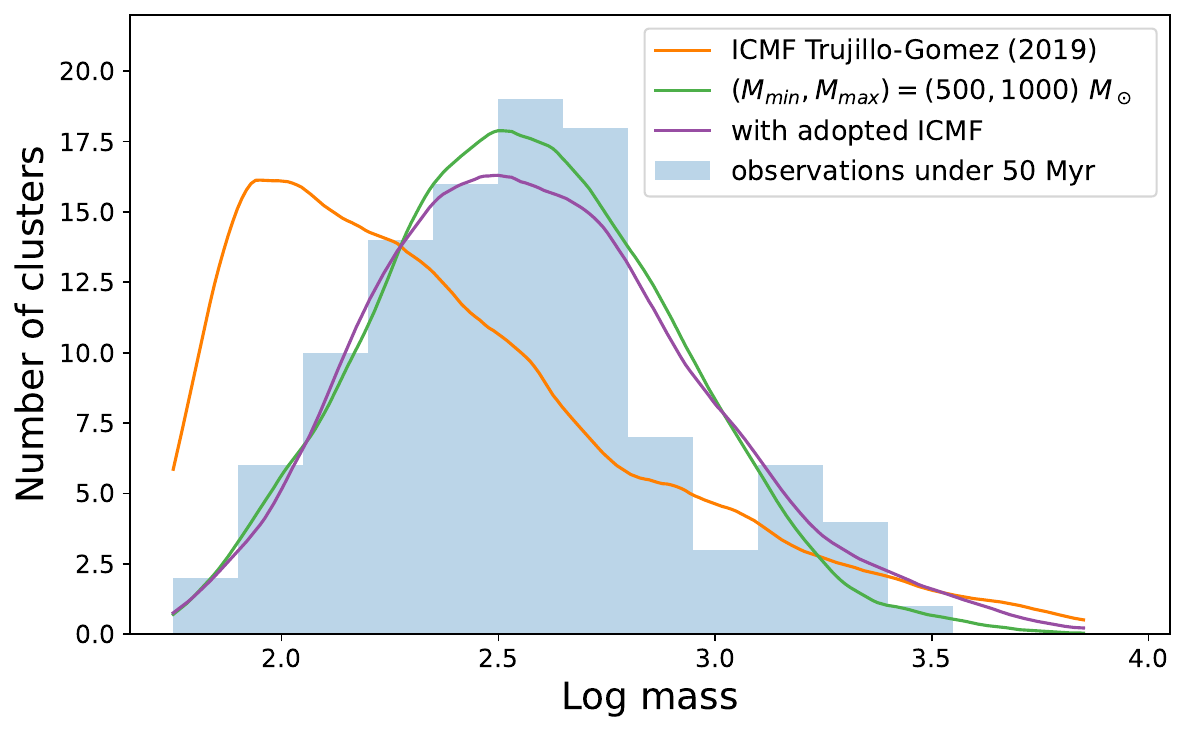}
    \caption{Distribution of the logarithm of mass for OCs with age under 50 Myr with simulated distribution of masses using different ICMFs: \citet{trujillo_2019}, \citet{trujillo_2019} with $M_{min} \sim$ 500 $M_\odot$ and $M_{max} \sim$ 1000 $M_\odot$, and the adopted ICMF (skewed log-normal with $\alpha$ = 2, location = 2.3 and scale = 0.5).}
    \label{fig:ICMF_fit}
\end{figure}

So, we proceed to estimate the $M_{\text{min}}$, and $M_{\text{max}}$ parameters of the ICMF using our sample of masses.
Since our sample includes few very young clusters, we use clusters younger than 50 Myr to have enough data (108 OCs in the silver sample) to make a meaningful analysis. Because at that age, clusters have already lost some mass we cannot directly determine the ICMF from the observed distribution. To take this into account, we perform a grid search of the optimal $M_{\text{min}}$, and $M_{\text{max}}$, in several runs of the model until 50 Myr. We used $t_4^{\text{tot}}$ = 2 Gyr and $\gamma$ = 0.6, which provided good fits for the age distribution. The best fits correspond to $M_{min} \sim$ 500 $M_\odot$ and $M_{max} \sim$ 1000 $M_\odot$, where a variation of about 20\% for these parameters still results in a mass distribution that is compatible with the observations. 
The fit is presented in Fig.~\ref{fig:ICMF_fit}, where the original ICMF from  \citet{trujillo_2019} is also presented for comparison. With the optimal parameters, the functional has a shape that resembles more an asymmetric log-normal with a turnover at the typical clusters mass $\sim$ 500 $M_\odot$. In this sense, the turnover parameter ($M_{min}$) does not define a minimum cluster mass, but a typical mass instead.

The truncated power-law ICMF with  $M_{min} \sim$ 500 $M_\odot$ and $M_{max} \sim$ 1000 $M_\odot$ provides a much better fit to the mass distribution of young clusters. However, we found that the number of high mass clusters was underestimated, and the low mass clusters over-estimated. This is not totally unexpected given the asymmetry apparent in Fig.~\ref{fig:ICMF_fit}.

To take this asymmetry into account, we tested a skew log-normal distribution, which has three parameters:  skewness $\alpha$, location (which is the mean when $\alpha = 0$) and scale \citep[standard deviation when $\alpha = 0$. See ][]{skew_normal_2009}. We used the SciPy $stats.skewnorm$ function to draw clusters masses. We found that the best fit was obtained with $\alpha$ = 2, location = 2.3 and scale = 0.5. As seen in Fig. \ref{fig:ICMF_fit}, it is broadly similar to the truncated-power law, also with a mode $\sim 500$ $M_{\sun}$, but produces a higher number of high mass clusters. We shall refer refer to this skewed log-normal ICMF as the "adopted ICMF", which will be used in the next section.

\subsection{Results with the adopted ICMF}

\begin{figure}[ht]
    \centering
    \includegraphics[width=0.9\linewidth]{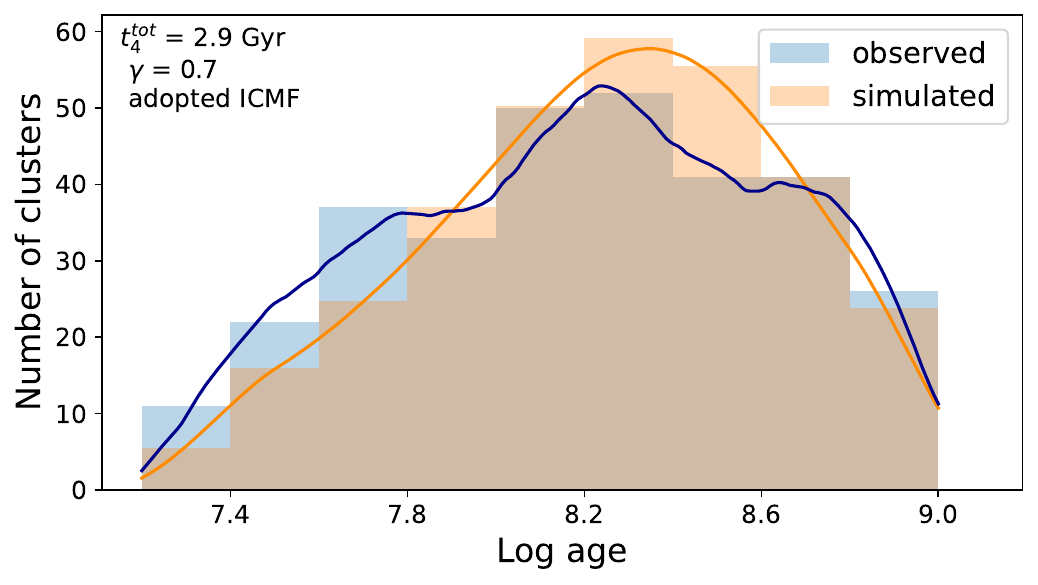}\
    \includegraphics[width=0.9\linewidth]{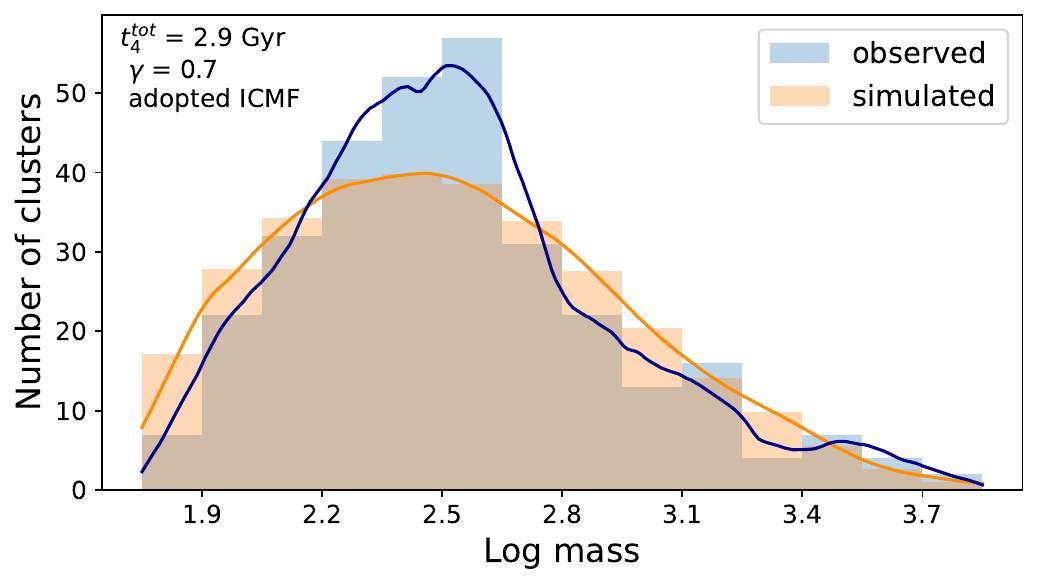}
    \caption{Comparison of the distribution of age (top) and mass (bottom) between the simulations (orange) and the observations (blue) considering $t_4^{\text{tot}}$ = 2.9 Gyr and $\gamma$ = 0.7, using the adopted ICMF. These are the optimal values found in this study.}
    \label{fig:DIST_silver_2kpc_mod_2}
\end{figure}

We ran the same grid as before for a first estimate the optimal values for $t_4^{\text{tot}}$ and $\gamma$ using the adopted ICMF. We find a good agreement with the observations for ($t_4^{\text{tot}}$, $\gamma$) = (3, 0.7); (3, 0.8) and (4, 0.8).
We thus ran a thinner grid with 0.1 Gyr steps in $t_4^{\text{tot}}$ and 0.05 steps in $\gamma$, and find the optimal value at $t_4^{\text{tot}} = 2.9 \pm 0.4$ Gyr and $\gamma = 0.7$. The uncertainty is taken as the standard deviation of 10 bootstrap samples. The results are represented in Fig.~\ref{fig:DIST_silver_2kpc_mod_2}. It can be seen that the model successfully reproduces the general characteristics of the age and mass distributions, such as the location of the peaks and broadness. However, the observed number of intermediate mass clusters ($\sim 500-600 M_{\sun}$) is higher.  At this stage, we could not determine the cause of this difference, whether it is due to the selection function of the data, or a missing ingredient in the model. Although we feel inclined towards the first possibility. This is a matter we are now investigating and will be presented in a follow-up work. A check using OCs within a shorter distance, under 1.5 kpc, shows similar trends with the best values found for $t_4^{\text{tot}}$ close to 3 Gyr and $\gamma$ around 0.6-0.7.

In any case, two results emerge from this analysis. The first is a strong indication that the initial mass distributions for embedded clusters and open clusters are substantially different. Qualitatively, the ICMF is bell shaped, which suggests a mass-independent star formation efficiency of the order of 20\% or lower \citep{parmentier_2008_icmf}. The second result, $t_4^{\text{tot}} = 2.9 \pm 0.4$ Gyr, is about twice the value of previous estimates, which were based on a power-law ICMF and used only the age distribution of OCs \citep{lamers_analytical_2005, anders_star_2021}.  This brings the disruption timescale closer to the larger disruption times found in n-body simulations \citep{baumgardt_dynamical_2003, PortegiesZwart_dissolution_1998}. This suggests that the disruption due to interactions with GMCs and spiral arms, while still the dominant channel of cluster disruption, may not be as strong as previously thought \citep{lamers_clusters_2006}.

\section{Conclusions \label{ch:conclusion}}

In this study, we have built a \textit{Gaia}-based catalogue of luminous masses for a sample of 1724 open clusters in the solar neighbourhood by comparing their luminosity distributions to theoretical luminosity functions. Our luminous mass distribution peaks at log(M) = 2.7 dex. 

Determining the masses, required the previous step of determining the cluster radii. The King ("limiting") and core radii were determined by fitting the King density profile to the observed profile of each cluster. The comparison between \citet{tarricq_structural_2022}, \citet{hunt_III_masses} and \citet{Just_2023} catalogues illustrates how hard the determination of cluster King limiting radii is, with large differences of the individual radii between catalogues. Nevertheless, our analysis shows that the uncertainties for the King radii do not have a significant impact on our luminous mass determinations.

Our analysis revealed a general agreement with the photometric masses from \citetalias{anderson_2023} and \citet{hunt_III_masses}, with a trend for higher masses in \citet{hunt_III_masses} which could be due to over-correction for Gaia incompleteness. In contrast, significant discrepancies are observed when compared with the (tidal) masses from \citet{Just_2023} which are highly affected by uncertainties in the tidal radius. Given these findings, we deem the luminous masses more reliable.

To study the mass loss rate in open clusters, we simulated the build up and mass evolution of a population of clusters following \citet{lamers_clusters_2006}. The model has three main ingredients: a cluster formation rate, which was assumed constant, an ICMF and disruption parameters ($t_4^{\text{tot}}$, and $\gamma$). 
Using the widely employed power-law ICMF, we obtain good fits to the age distribution, as was found in previous studies \citep[e.g.][]{lamers_clusters_2006, anders_star_2021}, although with longer dissolution times. However, the simulated mass distribution does not agree with the observations with any combination of parameters. As our model starts from the moment clusters emerge from the parent molecular clouds, we conjecture that the power-law distribution which had been determined for embedded clusters, might not be a good description of the initial masses of bound open clusters. 

Following previous indications that the ICMF may be bell shaped \citep{parmentier_2008_icmf, trujillo_2019}, we find that a skew log-normal ICMF provides a good match to the observations. The difference with respect to a power-law distribution of the embedded counterparts could indicate a typical star formation efficiency of $\leq 20\%$ in solar neighbourhood clusters \citep{parmentier_2008_icmf}. Finally, we find indications of a lower limit of $\sim 60 M{_\odot}$ for bound stellar clusters in the solar neighbourhood.

We note two caveats in this study: 
(1) Systematic uncertainties in the mass determinations could lead to overestimations of OC masses by up to 20\%, primarily due to the escape of low-mass stars through mass-segregated evaporation. The exact impact depends on the cluster's dynamical state. Correcting for this effect would likely reduce the value of $t_4$. However, additional simulations indicate that this reduction would not lower $t_4$ below $\sim 2.4$ Gyr.
(2) Variations in binary star fractions and mass ratios within the tidal radius could further affect the mass estimates.

Some questions also remain open, most noticeably, an excess of intermediate mass OCs that are not explained by the model. This may be due to data selection effects. In an era of continuous reports of new OC discoveries, the assessment of the OC selection function, and its dependence on distance, age and mass is badly needed. For this, stringent quality control of the growing sample of OCs and their memberships is essential. Finally, we note that more robust determinations of OC limiting radii will be crucial for determining dynamical based masses for comparison. Addressing these caveats and open questions will be the focus of our following studies.

\section*{Data availability}

Table \ref{tab:OC_results} is available in electronic form at the CDS via anonymous ftp to \href{http://cdsarc.u-strasbg.fr/}{cdsarc.u-strasbg.fr} (130.79.128.5) or via \href{http://cdsweb.u-strasbg.fr/cgi-bin/qcat?J/A+A/}{http://cdsweb.u-strasbg.fr/cgi-bin/qcat?J/A+A/}.

\begin{acknowledgements} 
We thank the anonymous referee for their constructive comments which have contributed to improve this paper. This work was partially supported by the Portuguese Fundação para a Ciência e a Tecnologia (FCT) through the Portuguese Strategic Programmes UIDB/FIS/00099/2020 and UIDP/FIS/00099/2020 for CENTRA. DA acknowledges support from the FCT/CENTRA PhD scholarship UI/BD/154465/2022. This research has made use of data from the European Space Agency (ESA) mission Gaia, processed by the Gaia Data Processing and Analysis Consortium (DPAC). We also thank Lorenzo Cavallo for insightful comments and suggestions.
\end{acknowledgements}

\bibliographystyle{aa} 
\bibliography{ocmass}

\begin{thebibliography}{73}
\expandafter\ifx\csname natexlab\endcsname\relax\def\natexlab#1{#1}\fi

\bibitem[{{Albrow}(2024)}]{2024MNRAS.528.6211A}
{Albrow}, M.~D. 2024, \mnras, 528, 6211

\bibitem[{Almeida {et~al.}(2023)Almeida, Monteiro, \& Dias}]{anderson_2023}
Almeida, A., Monteiro, H., \& Dias, W.~S. 2023, \mnras, 525, 2315

\bibitem[{Anders {et~al.}(2021)Anders, Cantat-Gaudin, Quadrino-Lodoso, Gieles,
  Jordi, Castro-Ginard, \& Balaguer-Núñez}]{anders_star_2021}
Anders, F., Cantat-Gaudin, T., Quadrino-Lodoso, I., {et~al.} 2021, \aap, 645,
  L2

\bibitem[{Azzalini \& Capitanio(2002)}]{skew_normal_2009}
Azzalini, A. \& Capitanio, A. 2002, Journal of the Royal Statistical Society
  Series B: Statistical Methodology, 61, 579

\bibitem[{{Baumgardt} \& {Kroupa}(2007)}]{BaumgardtKroupa2007}
{Baumgardt}, H. \& {Kroupa}, P. 2007, \mnras, 380, 1589

\bibitem[{Baumgardt \& Makino(2003)}]{baumgardt_dynamical_2003}
Baumgardt, H. \& Makino, J. 2003, \mnras, 340, 227

\bibitem[{{Bergond} {et~al.}(2001){Bergond}, {Leon}, \&
  {Guibert}}]{2001A&A...377..462B}
{Bergond}, G., {Leon}, S., \& {Guibert}, J. 2001, \aap, 377, 462

\bibitem[{Bossini {et~al.}(2019)Bossini, Vallenari, Bragaglia, Cantat-Gaudin,
  Sordo, Balaguer-Núñez, Jordi, Moitinho, Soubiran, Casamiquela, Carrera, \&
  Heiter}]{bossini_age_2019}
Bossini, D., Vallenari, A., Bragaglia, A., {et~al.} 2019, \aap, 623, A108

\bibitem[{Boutloukos \& Lamers(2003)}]{boutloukos_star_2003}
Boutloukos, S.~G. \& Lamers, H. 2003, \mnras, 338, 717

\bibitem[{{Bressan} {et~al.}(2012){Bressan}, {Marigo}, {Girardi}, {Salasnich},
  {Dal Cero}, {Rubele}, \& {Nanni}}]{2012MNRAS.427..127B}
{Bressan}, A., {Marigo}, P., {Girardi}, L., {et~al.} 2012, \mnras, 427, 127

\bibitem[{Cantat-Gaudin {et~al.}(2020)Cantat-Gaudin, Anders, Castro-Ginard,
  Jordi, Romero-Gómez, Soubiran, Casamiquela, Tarricq, Moitinho, Vallenari,
  Bragaglia, Krone-Martins, \& Kounkel}]{cantat-gaudin_painting_2020}
Cantat-Gaudin, T., Anders, F., Castro-Ginard, A., {et~al.} 2020, \aap, 640, A1

\bibitem[{Cantat-Gaudin {et~al.}(2018)Cantat-Gaudin, Jordi, Vallenari,
  Bragaglia, Balaguer-Núñez, Soubiran, Bossini, Moitinho, Castro-Ginard,
  Krone-Martins, Casamiquela, Sordo, \& Carrera}]{cantat-gaudin_gaia_2018}
Cantat-Gaudin, T., Jordi, C., Vallenari, A., {et~al.} 2018, \aap, 618, A93

\bibitem[{{Cantat-Gaudin} {et~al.}(2019){Cantat-Gaudin}, {Krone-Martins},
  {Sedaghat}, {Farahi}, {de Souza}, {Skalidis}, {Malz}, {Mac{\^e}do}, {Moews},
  {Jordi}, {Moitinho}, {Castro-Ginard}, {Ishida}, {Heneka}, {Boucaud}, \&
  {Trindade}}]{cantat_gaudin_incomplete_2019}
{Cantat-Gaudin}, T., {Krone-Martins}, A., {Sedaghat}, N., {et~al.} 2019, \aap,
  624, A126

\bibitem[{Castro-Ginard {et~al.}(2022)Castro-Ginard, Jordi, Luri,
  Cantat-Gaudin, Carrasco, Casamiquela, Anders, Balaguer-Núñez, \&
  Badia}]{castro-ginard_hunting_2022}
Castro-Ginard, A., Jordi, C., Luri, X., {et~al.} 2022, \aap, 661, A118

\bibitem[{{Castro-Ginard} {et~al.}(2018){Castro-Ginard}, {Jordi}, {Luri},
  {Julbe}, {Morvan}, {Balaguer-N{\'u}{\~n}ez}, \&
  {Cantat-Gaudin}}]{castro_2018_nearby}
{Castro-Ginard}, A., {Jordi}, C., {Luri}, X., {et~al.} 2018, \aap, 618, A59

\bibitem[{Castro-Ginard {et~al.}(2020)Castro-Ginard, Jordi, Luri, Álvarez
  Cid-Fuentes, Casamiquela, Anders, Cantat-Gaudin, Monguió, Balaguer-Núñez,
  Solà, \& Badia}]{castro-ginard_hunting_2020}
Castro-Ginard, A., Jordi, C., Luri, X., {et~al.} 2020, \aap, 635, A45

\bibitem[{{Cavallo} {et~al.}(2024){Cavallo}, {Spina}, {Carraro}, {Magrini},
  {Poggio}, {Cantat-Gaudin}, {Pasquato}, {Lucatello}, {Ortolani}, \&
  {Schiappacasse-Ulloa}}]{lorenzo_quad_2024}
{Cavallo}, L., {Spina}, L., {Carraro}, G., {et~al.} 2024, \aj, 167, 12

\bibitem[{{Chabrier}(2003)}]{Chabrier_2003}
{Chabrier}, G. 2003, \pasp, 115, 763

\bibitem[{{Chen} {et~al.}(2015){Chen}, {Bressan}, {Girardi}, {Marigo}, {Kong},
  \& {Lanza}}]{chen_2015}
{Chen}, Y., {Bressan}, A., {Girardi}, L., {et~al.} 2015, \mnras, 452, 1068

\bibitem[{{Cordoni} {et~al.}(2023){Cordoni}, {Milone}, {Marino}, {Vesperini},
  {Dondoglio}, {Legnardi}, {Mohandasan}, {Carlos}, {Lagioia}, {Jang}, \&
  {Ziliotto}}]{Cordoni_2023}
{Cordoni}, G., {Milone}, A.~P., {Marino}, A.~F., {et~al.} 2023, \aap, 672, A29

\bibitem[{{Dalessandro} {et~al.}(2015){Dalessandro}, {Miocchi}, {Carraro},
  {J{\'\i}lkov{\'a}}, \& {Moitinho}}]{2015MNRAS.449.1811D}
{Dalessandro}, E., {Miocchi}, P., {Carraro}, G., {J{\'\i}lkov{\'a}}, L., \&
  {Moitinho}, A. 2015, \mnras, 449, 1811

\bibitem[{{Della Croce} {et~al.}(2024){Della Croce}, {Dalessandro},
  {Livernois}, \& {Vesperini}}]{2023arXiv231202263D}
{Della Croce}, A., {Dalessandro}, E., {Livernois}, A., \& {Vesperini}, E. 2024,
  \aap, 683, A10

\bibitem[{{Dias} \& {L{\'e}pine}(2005)}]{dias_lepine_2005}
{Dias}, W.~S. \& {L{\'e}pine}, J.~R.~D. 2005, \apj, 629, 825

\bibitem[{Dias {et~al.}(2021)Dias, Monteiro, Moitinho, Lépine, Carraro,
  Paunzen, Alessi, \& Villela}]{dias_updated_2021}
Dias, W.~S., Monteiro, H., Moitinho, A., {et~al.} 2021, \mnras, 504, 356

\bibitem[{Ferreira {et~al.}(2021)Ferreira, Corradi, Maia, Angelo, \&
  Santos}]{ferreira_new_2021}
Ferreira, F.~A., Corradi, W. J.~B., Maia, F. F.~S., Angelo, M.~S., \& Santos,
  Jr., J. F.~C. 2021, \mnras, 502, L90

\bibitem[{Foreman-Mackey {et~al.}(2013)Foreman-Mackey, Hogg, Lang, \&
  Goodman}]{foreman-mackey_emcee_2013}
Foreman-Mackey, D., Hogg, D.~W., Lang, D., \& Goodman, J. 2013, \pasp, 125, 306

\bibitem[{{Gaia Collaboration} {et~al.}(2018){Gaia Collaboration}, {Brown},
  {Vallenari}, {Prusti}, {de Bruijne}, {Babusiaux}, {Bailer-Jones}, {Biermann},
  {Evans}, {Eyer}, {Jansen}, {Jordi}, {Klioner}, {Lammers}, {Lindegren},
  {Luri}, {Mignard}, {Panem}, {Pourbaix}, {Randich}, {Sartoretti}, {Siddiqui},
  {Soubiran}, {van Leeuwen}, {Walton}, {Arenou}, {Bastian}, {Cropper},
  {Drimmel}, {Katz}, {Lattanzi}, {Bakker}, {Cacciari}, {Casta{\~n}eda},
  {Chaoul}, {Cheek}, {De Angeli}, {Fabricius}, {Guerra}, {Holl}, {Masana},
  {Messineo}, {Mowlavi}, {Nienartowicz}, {Panuzzo}, {Portell}, {Riello},
  {Seabroke}, {Tanga}, {Th{\'e}venin}, {Gracia-Abril}, {Comoretto},
  {Garcia-Reinaldos}, {Teyssier}, {Altmann}, {Andrae}, {Audard},
  {Bellas-Velidis}, {Benson}, {Berthier}, {Blomme}, {Burgess}, {Busso},
  {Carry}, {Cellino}, {Clementini}, {Clotet}, {Creevey}, {Davidson}, {De
  Ridder}, {Delchambre}, {Dell'Oro}, {Ducourant},
  {Fern{\'a}ndez-Hern{\'a}ndez}, {Fouesneau}, {Fr{\'e}mat}, {Galluccio},
  {Garc{\'\i}a-Torres}, {Gonz{\'a}lez-N{\'u}{\~n}ez}, {Gonz{\'a}lez-Vidal},
  {Gosset}, {Guy}, {Halbwachs}, {Hambly}, {Harrison}, {Hern{\'a}ndez},
  {Hestroffer}, {Hodgkin}, {Hutton}, {Jasniewicz}, {Jean-Antoine-Piccolo},
  {Jordan}, {Korn}, {Krone-Martins}, {Lanzafame}, {Lebzelter}, {L{\"o}ffler},
  {Manteiga}, {Marrese}, {Mart{\'\i}n-Fleitas}, {Moitinho}, {Mora}, {Muinonen},
  {Osinde}, {Pancino}, {Pauwels}, {Petit}, {Recio-Blanco}, {Richards},
  {Rimoldini}, {Robin}, {Sarro}, {Siopis}, {Smith}, {Sozzetti}, {S{\"u}veges},
  {Torra}, {van Reeven}, {Abbas}, {Abreu Aramburu}, {Accart}, {Aerts},
  {Altavilla}, {{\'A}lvarez}, {Alvarez}, {Alves}, {Anderson}, {Andrei},
  {Anglada Varela}, {Antiche}, {Antoja}, {Arcay}, {Astraatmadja}, {Bach},
  {Baker}, {Balaguer-N{\'u}{\~n}ez}, {Balm}, {Barache}, {Barata}, {Barbato},
  {Barblan}, {Barklem}, {Barrado}, {Barros}, {Barstow}, {Bartholom{\'e}
  Mu{\~n}oz}, {Bassilana}, {Becciani}, {Bellazzini}, {Berihuete}, {Bertone},
  {Bianchi}, {Bienaym{\'e}}, {Blanco-Cuaresma}, {Boch}, {Boeche}, {Bombrun},
  {Borrachero}, {Bossini}, {Bouquillon}, {Bourda}, {Bragaglia}, {Bramante},
  {Breddels}, {Bressan}, {Brouillet}, {Br{\"u}semeister}, {Brugaletta},
  {Bucciarelli}, {Burlacu}, {Busonero}, {Butkevich}, {Buzzi}, {Caffau},
  {Cancelliere}, {Cannizzaro}, {Cantat-Gaudin}, {Carballo}, {Carlucci},
  {Carrasco}, {Casamiquela}, {Castellani}, {Castro-Ginard}, {Charlot},
  {Chemin}, {Chiavassa}, {Cocozza}, {Costigan}, {Cowell}, {Crifo}, {Crosta},
  {Crowley}, {Cuypers}, {Dafonte}, {Damerdji}, {Dapergolas}, {David}, {David},
  {de Laverny}, {De Luise}, {De March}, {de Martino}, {de Souza}, {de Torres},
  {Debosscher}, {del Pozo}, {Delbo}, {Delgado}, {Delgado}, {Valentini},
  {Valette}, {van Elteren}, {Van Hemelryck}, {van Leeuwen}, {Vaschetto},
  {Vecchiato}, {Veljanoski}, {Viala}, {Vicente}, {Vogt}, {von Essen}, {Voss},
  {Votruba}, {Voutsinas}, {Walmsley}, {Weiler}, {Wertz}, {Wevers},
  {Wyrzykowski}, {Yoldas}, {{\v{Z}}erjal}, {Ziaeepour}, {Zorec}, {Zschocke},
  {Zucker}, {Zurbach}, \& {Zwitter}}]{gaia_2_2018}
{Gaia Collaboration}, {Brown}, A.~G.~A., {Vallenari}, A., {et~al.} 2018, \aap,
  616, A1

\bibitem[{{Gaia Collaboration} {et~al.}(2021){Gaia Collaboration}, {Brown},
  {Vallenari}, {Prusti}, {de Bruijne}, {Babusiaux}, {Biermann}, {Creevey},
  {Evans}, {Eyer}, {Hutton}, {Jansen}, {Jordi}, {Klioner}, {Lammers},
  {Lindegren}, {Luri}, {Mignard}, {Panem}, {Pourbaix}, {Randich}, {Sartoretti},
  {Soubiran}, {Walton}, {Arenou}, {Bailer-Jones}, {Bastian}, {Cropper},
  {Drimmel}, {Katz}, {Lattanzi}, {van Leeuwen}, {Bakker}, {Cacciari},
  {Casta{\~n}eda}, {De Angeli}, {Ducourant}, {Fabricius}, {Fouesneau},
  {Fr{\'e}mat}, {Guerra}, {Guerrier}, {Guiraud}, {Jean-Antoine Piccolo},
  {Masana}, {Messineo}, {Mowlavi}, {Nicolas}, {Nienartowicz}, {Pailler},
  {Panuzzo}, {Riclet}, {Roux}, {Seabroke}, {Sordo}, {Tanga}, {Th{\'e}venin},
  {Gracia-Abril}, {Portell}, {Teyssier}, {Altmann}, {Andrae}, {Bellas-Velidis},
  {Benson}, {Berthier}, {Blomme}, {Brugaletta}, {Burgess}, {Busso}, {Carry},
  {Cellino}, {Cheek}, {Clementini}, {Damerdji}, {Davidson}, {Delchambre},
  {Dell'Oro}, {Fern{\'a}ndez-Hern{\'a}ndez}, {Galluccio}, {Garc{\'\i}a-Lario},
  {Garcia-Reinaldos}, {Gonz{\'a}lez-N{\'u}{\~n}ez}, {Gosset}, {Haigron},
  {Halbwachs}, {Hambly}, {Harrison}, {Hatzidimitriou}, {Heiter},
  {Hern{\'a}ndez}, {Hestroffer}, {Hodgkin}, {Holl}, {Jan{\ss}en}, {Jevardat de
  Fombelle}, {Jordan}, {Krone-Martins}, {Lanzafame}, {L{\"o}ffler}, {Lorca},
  {Manteiga}, {Marchal}, {Marrese}, {Moitinho}, {Mora}, {Muinonen}, {Osborne},
  {Pancino}, {Pauwels}, {Petit}, {Recio-Blanco}, {Richards}, {Riello},
  {Rimoldini}, {Robin}, {Roegiers}, {Rybizki}, {Sarro}, {Siopis}, {Smith},
  {Sozzetti}, {Ulla}, {Utrilla}, {van Leeuwen}, {van Reeven}, {Abbas}, {Abreu
  Aramburu}, {Accart}, {Aerts}, {Aguado}, {Ajaj}, {Altavilla}, {{\'A}lvarez},
  {{\'A}lvarez Cid-Fuentes}, {Alves}, {Anderson}, {Anglada Varela}, {Antoja},
  {Audard}, {Baines}, {Baker}, {Balaguer-N{\'u}{\~n}ez}, {Balbinot}, {Balog},
  {Barache}, {Barbato}, {Barros}, {Barstow}, {Bartolom{\'e}}, {Bassilana},
  {Bauchet}, {Baudesson-Stella}, {Becciani}, {Bellazzini}, {Bernet}, {Bertone},
  {Bianchi}, {Blanco-Cuaresma}, {Boch}, {Bombrun}, {Bossini}, {Bouquillon},
  {Bragaglia}, {Bramante}, {Breedt}, {Bressan}, {Brouillet}, {Bucciarelli},
  {Burlacu}, {Busonero}, {Butkevich}, {Buzzi}, {Caffau}, {Cancelliere},
  {C{\'a}novas}, {Cantat-Gaudin}, {Carballo}, {Carlucci}, {Carnerero},
  {Carrasco}, {Casamiquela}, {Castellani}, {Castro-Ginard}, {Castro Sampol},
  {Chaoul}, {Charlot}, {Chemin}, {Chiavassa}, {Cioni}, {Comoretto}, {Cooper},
  {Cornez}, {Cowell}, {Crifo}, {Crosta}, {Crowley}, {Dafonte}, {Dapergolas},
  {David}, {David}, {de Laverny}, {De Luise}, {De March}, {De Ridder}, {de
  Souza}, {de Teodoro}, {de Torres}, {del Peloso}, {del Pozo}, {Delbo},
  {Delgado}, {Delgado}, {Delisle}, {Di Matteo}, {Diakite}, {Diener},
  {Distefano}, {Dolding}, {Eappachen}, {Edvardsson}, {Enke}, {Esquej}, {Fabre},
  {Fabrizio}, {Faigler}, {Fedorets}, {Fernique}, {Fienga}, {Figueras},
  {Fouron}, {Fragkoudi}, {Fraile}, {Franke}, {Gai}, {Garabato},
  {Garcia-Gutierrez}, {Garc{\'\i}a-Torres}, {Garofalo}, {Gavras}, {Gerlach},
  {Geyer}, {Giacobbe}, {Gilmore}, {Girona}, {Giuffrida}, {Gomel}, {Gomez},
  {Gonzalez-Santamaria}, {Gonz{\'a}lez-Vidal}, {Granvik},
  {Guti{\'e}rrez-S{\'a}nchez}, {Guy}, {Hauser}, {Haywood}, {Helmi}, {Hidalgo},
  {Hilger}, {H{\l}adczuk}, {Hobbs}, {Holland}, {Huckle}, {Jasniewicz},
  {Jonker}, {Juaristi Campillo}, {Julbe}, {Karbevska}, {Kervella}, {Khanna},
  {Kochoska}, {Kontizas}, {Kordopatis}, {Korn}, {Kostrzewa-Rutkowska},
  {Kruszy{\'n}ska}, {Lambert}, {Lanza}, {Lasne}, {Le Campion}, {Le Fustec},
  {Lebreton}, {Lebzelter}, {Leccia}, {Leclerc}, {Lecoeur-Taibi}, {Liao},
  {Licata}, {Lindstr{\o}m}, {Lister}, {Livanou}, {Lobel}, {Madrero Pardo},
  {Managau}, {Mann}, {Marchant}, {Marconi}, {Marcos Santos}, {Marinoni},
  {Marocco}, {Marshall}, {Martin Polo}, {Mart{\'\i}n-Fleitas}, {Masip},
  {Massari}, {Mastrobuono-Battisti}, {Mazeh}, {McMillan}, {Messina},
  {Michalik}, {Millar}, {Mints}, {Molina}, {Molinaro}, {Moln{\'a}r},
  {Montegriffo}, {Mor}, {Morbidelli}, {Morel}, {Morris}, {Mulone}, {Munoz},
  {Muraveva}, {Murphy}, {Musella}, {Noval}, {Ord{\'e}novic}, {Orr{\`u}},
  {Osinde}, {Pagani}, {Pagano}, {Palaversa}, {Palicio}, {Panahi}, {Pawlak},
  {Pe{\~n}alosa Esteller}, {Penttil{\"a}}, {Piersimoni}, {Pineau}, {Plachy},
  {Plum}, {Poggio}, {Poretti}, {Poujoulet}, {Pr{\v{s}}a}, {Pulone}, {Racero},
  {Ragaini}, {Rainer}, {Raiteri}, {Rambaux}, {Ramos}, {Ramos-Lerate}, {Re
  Fiorentin}, {Regibo}, {Reyl{\'e}}, {Ripepi}, {Riva}, {Rixon}, {Robichon},
  {Robin}, {Roelens}, {Rohrbasser}, {Romero-G{\'o}mez}, {Rowell}, {Royer},
  {Rybicki}, {Sadowski}, {Sagrist{\`a} Sell{\'e}s}, {Sahlmann}, {Salgado},
  {Salguero}, {Samaras}, {Sanchez Gimenez}, {Sanna}, {Santove{\~n}a},
  {Sarasso}, {Schultheis}, {Sciacca}, {Segol}, {Segovia}, {S{\'e}gransan},
  {Semeux}, {Shahaf}, {Siddiqui}, {Siebert}, {Siltala}, {Slezak}, {Smart},
  {Solano}, {Solitro}, {Souami}, {Souchay}, {Spagna}, {Spoto}, {Steele},
  {Steidelm{\"u}ller}, {Stephenson}, {S{\"u}veges}, {Szabados}, {Szegedi-Elek},
  {Taris}, {Tauran}, {Taylor}, {Teixeira}, {Thuillot}, {Tonello}, {Torra},
  {Torra}, {Turon}, {Unger}, {Vaillant}, {van Dillen}, {Vanel}, {Vecchiato},
  {Viala}, {Vicente}, {Voutsinas}, {Weiler}, {Wevers}, {Wyrzykowski}, {Yoldas},
  {Yvard}, {Zhao}, {Zorec}, {Zucker}, {Zurbach}, \&
  {Zwitter}}]{2021A&A...649A...1G}
{Gaia Collaboration}, {Brown}, A.~G.~A., {Vallenari}, A., {et~al.} 2021, \aap,
  649, A1

\bibitem[{{Gaia Collaboration} {et~al.}(2016{\natexlab{a}}){Gaia
  Collaboration}, {Brown}, {Vallenari}, {Prusti}, {de Bruijne}, {Mignard},
  {Drimmel}, {Babusiaux}, {Bailer-Jones}, {Bastian}, {Biermann}, {Evans},
  {Eyer}, {Jansen}, {Jordi}, {Katz}, {Klioner}, {Lammers}, {Lindegren}, {Luri},
  {O'Mullane}, {Panem}, {Pourbaix}, {Randich}, {Sartoretti}, {Siddiqui},
  {Soubiran}, {Valette}, {van Leeuwen}, {Walton}, {Aerts}, {Arenou}, {Cropper},
  {H{\o}g}, {Lattanzi}, {Grebel}, {Holland}, {Huc}, {Passot}, {Perryman},
  {Bramante}, {Cacciari}, {Casta{\~n}eda}, {Chaoul}, {Cheek}, {De Angeli},
  {Fabricius}, {Guerra}, {Hern{\'a}ndez}, {Jean-Antoine-Piccolo}, {Masana},
  {Messineo}, {Mowlavi}, {Nienartowicz}, {Ord{\'o}{\~n}ez-Blanco}, {Panuzzo},
  {Portell}, {Richards}, {Riello}, {Seabroke}, {Tanga}, {Th{\'e}venin},
  {Torra}, {Els}, {Gracia-Abril}, {Comoretto}, {Garcia-Reinaldos}, {Lock},
  {Mercier}, {Altmann}, {Andrae}, {Astraatmadja}, {Bellas-Velidis}, {Benson},
  {Berthier}, {Blomme}, {Busso}, {Carry}, {Cellino}, {Clementini}, {Cowell},
  {Creevey}, {Cuypers}, {Davidson}, {De Ridder}, {de Torres}, {Delchambre},
  {Dell'Oro}, {Ducourant}, {Fr{\'e}mat}, {Garc{\'\i}a-Torres}, {Gosset},
  {Halbwachs}, {Hambly}, {Harrison}, {Hauser}, {Hestroffer}, {Hodgkin},
  {Huckle}, {Hutton}, {Jasniewicz}, {Jordan}, {Kontizas}, {Korn}, {Lanzafame},
  {Manteiga}, {Moitinho}, {Muinonen}, {Osinde}, {Pancino}, {Pauwels}, {Petit},
  {Recio-Blanco}, {Robin}, {Sarro}, {Siopis}, {Smith}, {Smith}, {Sozzetti},
  {Thuillot}, {van Reeven}, {Viala}, {Abbas}, {Abreu Aramburu}, {Accart},
  {Aguado}, {Allan}, {Allasia}, {Altavilla}, {{\'A}lvarez}, {Alves},
  {Anderson}, {Andrei}, {Anglada Varela}, {Antiche}, {Antoja}, {Ant{\'o}n},
  {Arcay}, {Bach}, {Baker}, {Balaguer-N{\'u}{\~n}ez}, {Barache}, {Barata},
  {Barbier}, {Barblan}, {Barrado y Navascu{\'e}s}, {Barros}, {Barstow},
  {Becciani}, {Bellazzini}, {Bello Garc{\'\i}a}, {Belokurov}, {Bendjoya},
  {Berihuete}, {Bianchi}, {Bienaym{\'e}}, {Billebaud}, {Blagorodnova},
  {Blanco-Cuaresma}, {Boch}, {Bombrun}, {Borrachero}, {Bouquillon}, {Bourda},
  {Bouy}, {Bragaglia}, {Breddels}, {Brouillet}, {Br{\"u}semeister},
  {Bucciarelli}, {Burgess}, {Burgon}, {Burlacu}, {Busonero}, {Buzzi}, {Caffau},
  {Cambras}, {Campbell}, {Cancelliere}, {Cantat-Gaudin}, {Carlucci},
  {Carrasco}, {Castellani}, {Charlot}, {Charnas}, {Chiavassa}, {Clotet},
  {Cocozza}, {Collins}, {Costigan}, {Crifo}, {Cross}, {Crosta}, {Crowley},
  {Dafonte}, {Damerdji}, {Dapergolas}, {David}, {David}, {De Cat}, {de Felice},
  {de Laverny}, {De Luise}, {De March}, {de Martino}, {de Souza}, {Debosscher},
  {del Pozo}, {Delbo}, {Delgado}, {Delgado}, {Di Matteo}, {Diakite},
  {Distefano}, {Dolding}, {Dos Anjos}, {Drazinos}, {Duran}, {Dzigan},
  {Edvardsson}, {Enke}, {Evans}, {Eynard Bontemps}, {Fabre}, {Fabrizio},
  {Faigler}, {Falc{\~a}o}, {Farr{\`a}s Casas}, {Federici}, {Fedorets},
  {Fern{\'a}ndez-Hern{\'a}ndez}, {Fernique}, {Fienga}, {Figueras}, {Filippi},
  {Findeisen}, {Fonti}, {Fouesneau}, {Fraile}, {Fraser}, {Fuchs}, {Gai},
  {Galleti}, {Galluccio}, {Garabato}, {Garc{\'\i}a-Sedano}, {Garofalo},
  {Garralda}, {Gavras}, {Gerssen}, {Geyer}, {Gilmore}, {Girona}, {Giuffrida},
  {Gomes}, {Gonz{\'a}lez-Marcos}, {Gonz{\'a}lez-N{\'u}{\~n}ez},
  {Gonz{\'a}lez-Vidal}, {Granvik}, {Guerrier}, {Guillout}, {Guiraud},
  {G{\'u}rpide}, {Guti{\'e}rrez-S{\'a}nchez}, {Guy}, {Haigron},
  {Hatzidimitriou}, {Haywood}, {Heiter}, {Helmi}, {Hobbs}, {Hofmann}, {Holl},
  {Holland}, {Hunt}, {Hypki}, {Icardi}, {Irwin}, {Jevardat de Fombelle},
  {Jofr{\'e}}, {Jonker}, {Jorissen}, {Julbe}, {Karampelas}, {Kochoska},
  {Kohley}, {Kolenberg}, {Kontizas}, {Koposov}, {Kordopatis}, {Koubsky},
  {Krone-Martins}, {Kudryashova}, {Kull}, {Bachchan}, {Lacoste-Seris}, {Lanza},
  {Lavigne}, {Le Poncin-Lafitte}, {Lebreton}, {Lebzelter}, {Leccia}, {Leclerc},
  {Lecoeur-Taibi}, {Lemaitre}, {Lenhardt}, {Leroux}, {Liao}, {Licata},
  {Lindstr{\o}m}, {Lister}, {Livanou}, {Lobel}, {L{\"o}ffler}, {L{\'o}pez},
  {Lorenz}, {MacDonald}, {Magalh{\~a}es Fernandes}, {Managau}, {Mann},
  {Mantelet}, {Marchal}, {Marchant}, {Marconi}, {Marinoni}, {Marrese},
  {Marschalk{\'o}}, {Marshall}, {Mart{\'\i}n-Fleitas}, {Martino}, {Mary},
  {Matijevi{\v{c}}}, {Mazeh}, {McMillan}, {Messina}, {Michalik}, {Millar},
  {Miranda}, {Molina}, {Molinaro}, {Molinaro}, {Moln{\'a}r}, {Moniez},
  {Montegriffo}, {Mor}, {Mora}, {Morbidelli}, {Morel}, {Morgenthaler},
  {Morris}, {Mulone}, {Muraveva}, {Musella}, {Narbonne}, {Nelemans},
  {Nicastro}, {Noval}, {Ord{\'e}novic}, {Ordieres-Mer{\'e}}, {Osborne},
  {Pagani}, {Pagano}, {Pailler}, {Palacin}, {Palaversa}, {Parsons}, {Pecoraro},
  {Pedrosa}, {Pentik{\"a}inen}, {Pichon}, {Piersimoni}, {Pineau}, {Plachy},
  {Plum}, {Poujoulet}, {Pr{\v{s}}a}, {Pulone}, {Ragaini}, {Rago}, {Rambaux},
  {Ramos-Lerate}, {Ranalli}, {Rauw}, {Read}, {Regibo}, {Reyl{\'e}}, {Ribeiro},
  {Rimoldini}, {Ripepi}, {Riva}, {Rixon}, {Roelens}, {Romero-G{\'o}mez},
  {Rowell}, {Royer}, {Ruiz-Dern}, {Sadowski}, {Sagrist{\`a} Sell{\'e}s},
  {Sahlmann}, {Salgado}, {Salguero}, {Sarasso}, {Savietto}, {Schultheis},
  {Sciacca}, {Segol}, {Segovia}, {Segransan}, {Shih}, {Smareglia}, {Smart},
  {Solano}, {Solitro}, {Sordo}, {Soria Nieto}, {Souchay}, {Spagna}, {Spoto},
  {Stampa}, {Steele}, {Steidelm{\"u}ller}, {Stephenson}, {Stoev}, {Suess},
  {S{\"u}veges}, {Surdej}, {Szabados}, {Szegedi-Elek}, {Tapiador}, {Taris},
  {Tauran}, {Taylor}, {Teixeira}, {Terrett}, {Tingley}, {Trager}, {Turon},
  {Ulla}, {Utrilla}, {Valentini}, {van Elteren}, {Van Hemelryck}, {van
  Leeuwen}, {Varadi}, {Vecchiato}, {Veljanoski}, {Via}, {Vicente}, {Vogt},
  {Voss}, {Votruba}, {Voutsinas}, {Walmsley}, {Weiler}, {Weingrill}, {Wevers},
  {Wyrzykowski}, {Yoldas}, {{\v{Z}}erjal}, {Zucker}, {Zurbach}, {Zwitter},
  {Alecu}, {Allen}, {Allende Prieto}, {Amorim}, {Anglada-Escud{\'e}},
  {Arsenijevic}, {Azaz}, {Balm}, {Beck}, {Bernstein}, {Bigot}, {Bijaoui},
  {Blasco}, {Bonfigli}, {Bono}, {Boudreault}, {Bressan}, {Brown}, {Brunet},
  {Bunclark}, {Buonanno}, {Butkevich}, {Carret}, {Carrion}, {Chemin},
  {Ch{\'e}reau}, {Corcione}, {Darmigny}, {de Boer}, {de Teodoro}, {de Zeeuw},
  {Delle Luche}, {Domingues}, {Dubath}, {Fodor}, {Fr{\'e}zouls}, {Fries},
  {Fustes}, {Fyfe}, {Gallardo}, {Gallegos}, {Gardiol}, {Gebran}, {Gomboc},
  {G{\'o}mez}, {Grux}, {Gueguen}, {Heyrovsky}, {Hoar}, {Iannicola}, {Isasi
  Parache}, {Janotto}, {Joliet}, {Jonckheere}, {Keil}, {Kim}, {Klagyivik},
  {Klar}, {Knude}, {Kochukhov}, {Kolka}, {Kos}, {Kutka}, {Lainey}, {LeBouquin},
  {Liu}, {Loreggia}, {Makarov}, {Marseille}, {Martayan}, {Martinez-Rubi},
  {Massart}, {Meynadier}, {Mignot}, {Munari}, {Nguyen}, {Nordlander}, {Ocvirk},
  {O'Flaherty}, {Olias Sanz}, {Ortiz}, {Osorio}, {Oszkiewicz}, {Ouzounis},
  {Palmer}, {Park}, {Pasquato}, {Peltzer}, {Peralta}, {P{\'e}turaud},
  {Pieniluoma}, {Pigozzi}, {Poels}, {Prat}, {Prod'homme}, {Raison}, {Rebordao},
  {Risquez}, {Rocca-Volmerange}, {Rosen}, {Ruiz-Fuertes}, {Russo}, {Sembay},
  {Serraller Vizcaino}, {Short}, {Siebert}, {Silva}, {Sinachopoulos}, {Slezak},
  {Soffel}, {Sosnowska}, {Strai{\v{z}}ys}, {ter Linden}, {Terrell}, {Theil},
  {Tiede}, {Troisi}, {Tsalmantza}, {Tur}, {Vaccari}, {Vachier}, {Valles}, {Van
  Hamme}, {Veltz}, {Virtanen}, {Wallut}, {Wichmann}, {Wilkinson}, {Ziaeepour},
  \& {Zschocke}}]{2016A&A...595A...2G}
{Gaia Collaboration}, {Brown}, A.~G.~A., {Vallenari}, A., {et~al.}
  2016{\natexlab{a}}, \aap, 595, A2

\bibitem[{{Gaia Collaboration} {et~al.}(2016{\natexlab{b}}){Gaia
  Collaboration}, {Prusti}, {de Bruijne}, {Brown}, {Vallenari}, {Babusiaux},
  {Bailer-Jones}, {Bastian}, {Biermann}, {Evans}, {Eyer}, {Jansen}, {Jordi},
  {Klioner}, {Lammers}, {Lindegren}, {Luri}, {Mignard}, {Milligan}, {Panem},
  {Poinsignon}, {Pourbaix}, {Randich}, {Sarri}, {Sartoretti}, {Siddiqui},
  {Soubiran}, {Valette}, {van Leeuwen}, {Walton}, {Aerts}, \&
  {Arenou}}]{gaia_mission_ref}
{Gaia Collaboration}, {Prusti}, T., {de Bruijne}, J.~H.~J., {et~al.}
  2016{\natexlab{b}}, \aap, 595, A1

\bibitem[{{Goodman} \& {Weare}(2010)}]{goodman_2010}
{Goodman}, J. \& {Weare}, J. 2010, Communications in Applied Mathematics and
  Computational Science, 5, 65

\bibitem[{{Guilherme-Garcia} {et~al.}(2023){Guilherme-Garcia}, {Krone-Martins},
  \& {Moitinho}}]{2023A&A...673A.128G}
{Guilherme-Garcia}, P., {Krone-Martins}, A., \& {Moitinho}, A. 2023, \aap, 673,
  A128

\bibitem[{{Hunt} \& {Reffert}(2021)}]{Hunt_I_2021}
{Hunt}, E.~L. \& {Reffert}, S. 2021, \aap, 646, A104

\bibitem[{{Hunt} \& {Reffert}(2023)}]{Hunt_II_2023}
{Hunt}, E.~L. \& {Reffert}, S. 2023, \aap, 673, A114

\bibitem[{{Hunt} \& {Reffert}(2024)}]{hunt_III_masses}
{Hunt}, E.~L. \& {Reffert}, S. 2024, \aap, 686, A42

\bibitem[{{Just} {et~al.}(2023){Just}, {Piskunov}, {Klos}, {Kovaleva}, \&
  {Polyachenko}}]{Just_2023}
{Just}, A., {Piskunov}, A.~E., {Klos}, J.~H., {Kovaleva}, D.~A., \&
  {Polyachenko}, E.~V. 2023, \aap, 672, A187

\bibitem[{Kharchenko(2001)}]{kharchenko_all-sky_2001}
Kharchenko, N.~V. 2001, Kinematika i Fizika Nebesnykh Tel, 17, 409

\bibitem[{{Kharchenko} {et~al.}(2005){Kharchenko}, {Piskunov}, {R{\"o}ser},
  {Schilbach}, \& {Scholz}}]{Kharchenko_2005}
{Kharchenko}, N.~V., {Piskunov}, A.~E., {R{\"o}ser}, S., {Schilbach}, E., \&
  {Scholz}, R.~D. 2005, \aap, 438, 1163

\bibitem[{Kharchenko {et~al.}(2013)Kharchenko, Piskunov, Schilbach, Röser, \&
  Scholz}]{kharchenko_global_2013}
Kharchenko, N.~V., Piskunov, A.~E., Schilbach, E., Röser, S., \& Scholz, R.-D.
  2013, \aap, 558, A53

\bibitem[{King(1962)}]{king_structure_1962}
King, I. 1962, \aj, 67, 471

\bibitem[{Knuth(2013)}]{knuth_optimal_2006}
Knuth, K.~H. 2013, ArXiv e-prints [\eprint[arXiv]{physics/0605197}]

\bibitem[{{Kroupa}(2001)}]{kroupa_2001}
{Kroupa}, P. 2001, \mnras, 322, 231

\bibitem[{Küpper {et~al.}(2010)Küpper, Kroupa, Baumgardt, \&
  Heggie}]{kupper_peculiarities_2010}
Küpper, A. H.~W., Kroupa, P., Baumgardt, H., \& Heggie, D.~C. 2010, \mnras,
  407, 2241

\bibitem[{Lada \& Lada(2003)}]{lada_embedded_2003}
Lada, C.~J. \& Lada, E.~A. 2003, \araa, 41, 57

\bibitem[{Lamers {et~al.}(2010)Lamers, Baumgardt, \& Gieles}]{lamers_mass_2010}
Lamers, Baumgardt, H., \& Gieles. 2010, \mnras

\bibitem[{Lamers \& Gieles(2006)}]{lamers_clusters_2006}
Lamers, H. \& Gieles, M. 2006, \aap, 455, L17

\bibitem[{Lamers {et~al.}(2005{\natexlab{a}})Lamers, Gieles, Bastian,
  Baumgardt, Kharchenko, \& Portegies~Zwart}]{lamers_analytical_2005}
Lamers, H., Gieles, M., Bastian, N., {et~al.} 2005{\natexlab{a}}, \aap, 441,
  117

\bibitem[{Lamers {et~al.}(2005{\natexlab{b}})Lamers, Gieles, \&
  Portegies~Zwart}]{lamers_disruption_2005}
Lamers, H., Gieles, M., \& Portegies~Zwart, S.~F. 2005{\natexlab{b}}, \aap,
  429, 173

\bibitem[{Liu \& Pang(2019)}]{liu_catalog_2019}
Liu, L. \& Pang, X. 2019, \apjs, 245, 32

\bibitem[{{Ma{\'\i}z Apell{\'a}niz} \& {Weiler}(2018)}]{mariz_weiler_2018}
{Ma{\'\i}z Apell{\'a}niz}, J. \& {Weiler}, M. 2018, \aap, 619, A180

\bibitem[{{Marigo} {et~al.}(2013){Marigo}, {Bressan}, {Nanni}, {Girardi}, \&
  {Pumo}}]{marigo}
{Marigo}, P., {Bressan}, A., {Nanni}, A., {Girardi}, L., \& {Pumo}, M.~L. 2013,
  \mnras, 434, 488

\bibitem[{{Marigo} {et~al.}(2008){Marigo}, {Girardi}, {Bressan}, {Groenewegen},
  {Silva}, \& {Granato}}]{marigo_scale_rel}
{Marigo}, P., {Girardi}, L., {Bressan}, A., {et~al.} 2008, \aap, 482, 883

\bibitem[{{Meingast} {et~al.}(2021){Meingast}, {Alves}, \&
  {Rottensteiner}}]{2021A&A...645A..84M}
{Meingast}, S., {Alves}, J., \& {Rottensteiner}, A. 2021, \aap, 645, A84

\bibitem[{{Moitinho}(2010)}]{Moitinho_2010}
{Moitinho}, A. 2010, in Star Clusters: Basic Galactic Building Blocks
  Throughout Time and Space, ed. R.~{de Grijs} \& J.~R.~D. {L{\'e}pine}, Vol.
  266, 106--116

\bibitem[{Monteiro {et~al.}(2021)Monteiro, Barros, Dias, \&
  Lépine}]{monteiro_distribution_2021}
Monteiro, H., Barros, D.~A., Dias, W.~S., \& Lépine, J. R.~D. 2021, Frontiers
  in Astronomy and Space Sciences, 8

\bibitem[{Monteiro {et~al.}(2017)Monteiro, Dias, Hickel, \&
  Caetano}]{monteiro_opd_2017}
Monteiro, H., Dias, W.~S., Hickel, G.~R., \& Caetano, T.~C. 2017, \na, 51, 15

\bibitem[{Monteiro {et~al.}(2020)Monteiro, Dias, Moitinho, Cantat-Gaudin,
  Lépine, Carraro, \& Paunzen}]{monteiro_fundamental_2020}
Monteiro, H., Dias, W.~S., Moitinho, A., {et~al.} 2020, \mnras, 499, 1874

\bibitem[{{Moreira} {et~al.}(2024){Moreira}, {Moitinho}, {Silva}, \&
  {Almeida}}]{sandro_2024}
{Moreira}, S., {Moitinho}, A., {Silva}, A., \& {Almeida}, D. 2024, arXiv
  e-prints, arXiv:2406.14661, submitted to Astronomy \& Astrophysics

\bibitem[{Newville {et~al.}(2014)Newville, Stensitzki, Allen, \&
  Ingargiola}]{newville_matthew_2014_11813}
Newville, M., Stensitzki, T., Allen, D.~B., \& Ingargiola, A. 2014, {LMFIT:
  Non-Linear Least-Square Minimization and Curve-Fitting for Python}

\bibitem[{{Parmentier} {et~al.}(2008){Parmentier}, {Goodwin}, {Kroupa}, \&
  {Baumgardt}}]{parmentier_2008_icmf}
{Parmentier}, G., {Goodwin}, S.~P., {Kroupa}, P., \& {Baumgardt}, H. 2008,
  \apj, 678, 347

\bibitem[{Piskunov {et~al.}(2008{\natexlab{a}})Piskunov, Kharchenko, Schilbach,
  Röser, Scholz, \& Zinnecker}]{piskunov_initial_2008}
Piskunov, A.~E., Kharchenko, N.~V., Schilbach, E., {et~al.} 2008{\natexlab{a}},
  \aap, 487, 557, number: 2 Publisher: EDP Sciences

\bibitem[{Piskunov {et~al.}(2007)Piskunov, Schilbach, Kharchenko, Röser, \&
  Scholz}]{piskunov_towards_2007}
Piskunov, A.~E., Schilbach, E., Kharchenko, N.~V., Röser, S., \& Scholz, R.-D.
  2007, \aap, 468, 151

\bibitem[{Piskunov {et~al.}(2008{\natexlab{b}})Piskunov, Schilbach, Kharchenko,
  Röser, \& Scholz}]{piskunov_tidal_2008}
Piskunov, A.~E., Schilbach, E., Kharchenko, N.~V., Röser, S., \& Scholz, R.-D.
  2008{\natexlab{b}}, \aap, 477, 165

\bibitem[{{Portegies Zwart} {et~al.}(1998){Portegies Zwart}, {Hut}, {Makino},
  \& {McMillan}}]{PortegiesZwart_dissolution_1998}
{Portegies Zwart}, S.~F., {Hut}, P., {Makino}, J., \& {McMillan}, S. L.~W.
  1998, \aap, 337, 363

\bibitem[{{Schmeja} {et~al.}(2014){Schmeja}, {Kharchenko}, {Piskunov},
  {R{\"o}ser}, {Schilbach}, {Froebrich}, \& {Scholz}}]{Schmeja_2014}
{Schmeja}, S., {Kharchenko}, N.~V., {Piskunov}, A.~E., {et~al.} 2014, \aap,
  568, A51

\bibitem[{{Scholz} {et~al.}(2015){Scholz}, {Kharchenko}, {Piskunov},
  {R{\"o}ser}, \& {Schilbach}}]{Scholz_2015}
{Scholz}, R.~D., {Kharchenko}, N.~V., {Piskunov}, A.~E., {R{\"o}ser}, S., \&
  {Schilbach}, E. 2015, \aap, 581, A39

\bibitem[{{Shukirgaliyev} {et~al.}(2017){Shukirgaliyev}, {Parmentier},
  {Berczik}, \& {Just}}]{Shukirgaliyev_2017}
{Shukirgaliyev}, B., {Parmentier}, G., {Berczik}, P., \& {Just}, A. 2017, \aap,
  605, A119

\bibitem[{{Shukirgaliyev} {et~al.}(2019){Shukirgaliyev}, {Parmentier},
  {Berczik}, \& {Just}}]{Shukirgaliyev_2019}
{Shukirgaliyev}, B., {Parmentier}, G., {Berczik}, P., \& {Just}, A. 2019,
  \mnras, 486, 1045

\bibitem[{Sim {et~al.}(2019)Sim, Lee, Ann, \& Kim}]{sim_207_2019}
Sim, G., Lee, S.~H., Ann, H.~B., \& Kim, S. 2019, Journal of Korean
  Astronomical Society, 52, 145

\bibitem[{{Snaith} {et~al.}(2015){Snaith}, {Haywood}, {Di Matteo}, {Lehnert},
  {Combes}, {Katz}, \& {G{\'o}mez}}]{Snaith_2015_sfr}
{Snaith}, O., {Haywood}, M., {Di Matteo}, P., {et~al.} 2015, \aap, 578, A87

\bibitem[{Tarricq {et~al.}(2022)Tarricq, Soubiran, Casamiquela, Castro-Ginard,
  Olivares, Miret-Roig, \& Galli}]{tarricq_structural_2022}
Tarricq, Y., Soubiran, C., Casamiquela, L., {et~al.} 2022, \aap, 659, A59

\bibitem[{{Trujillo-Gomez} {et~al.}(2019){Trujillo-Gomez}, {Reina-Campos}, \&
  {Kruijssen}}]{trujillo_2019}
{Trujillo-Gomez}, S., {Reina-Campos}, M., \& {Kruijssen}, J.~M.~D. 2019,
  \mnras, 488, 3972

\bibitem[{Virtanen {et~al.}(2020)Virtanen, Gommers, Oliphant, Haberland, Reddy,
  Cournapeau, Burovski, Peterson, Weckesser, Bright, {van der Walt}, Brett,
  Wilson, Millman, Mayorov, Nelson, Jones, Kern, Larson, Carey, Polat, Feng,
  Moore, {VanderPlas}, Laxalde, Perktold, Cimrman, Henriksen, Quintero, Harris,
  Archibald, Ribeiro, Pedregosa, {van Mulbregt}, \& {SciPy 1.0
  Contributors}}]{2020SciPy-NMeth}
Virtanen, P., Gommers, R., Oliphant, T.~E., {et~al.} 2020, Nature Methods, 17,
  261

\end{thebibliography}

\begin{appendix}

\section{Distribution of \texorpdfstring{$N_0$}{N0} and c \label{ch:appendixA}}

\begin{figure}[ht]
    \centering
    \includegraphics[width=0.9\linewidth]{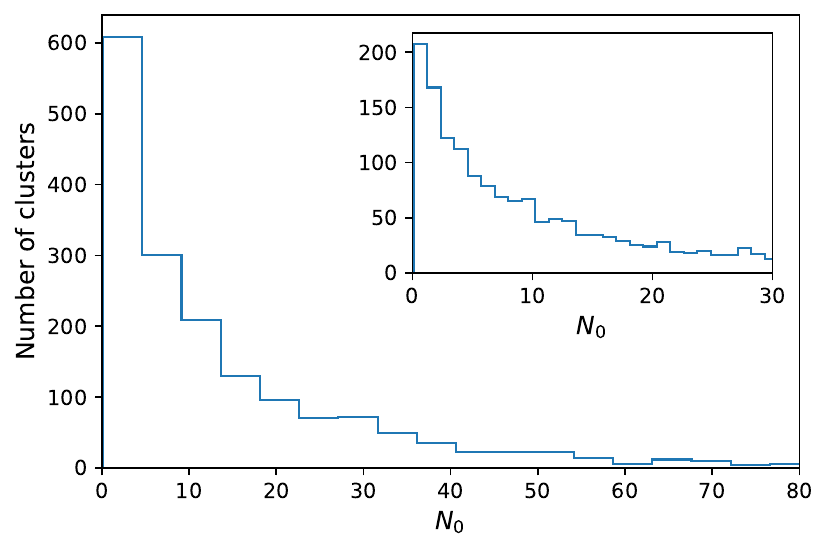}
    \caption{Distribution of parameter $N_0$. There are 35 OCs with $N_0$ above 80 which were not displayed to allow an easier visualization.}
    \label{fig:n0}
\end{figure}

\begin{figure}[ht]
    \centering
    \includegraphics[width=0.9\linewidth]{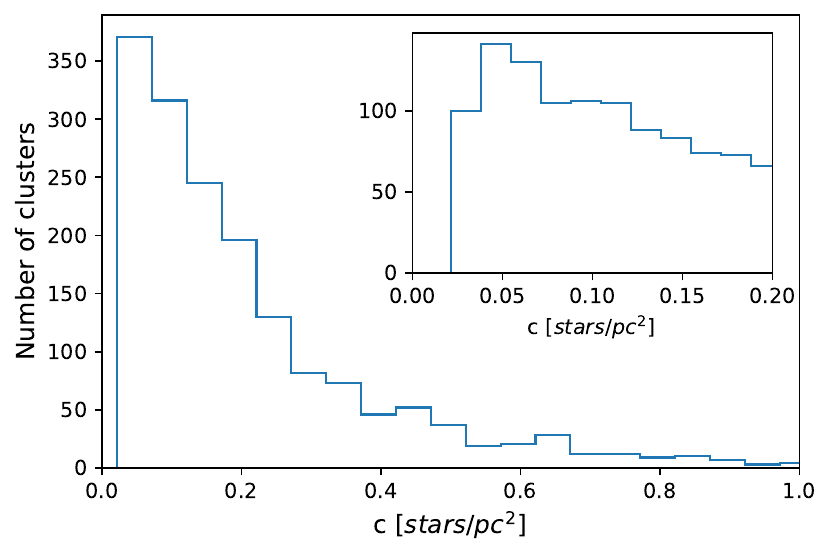}
    \caption{Distribution of parameter c. There are 53 OCs with c above 1 $star/pc^2$ which were not displayed to allow an easier visualization.}
    \label{fig:c}
\end{figure}

\newpage

\section{Comparison of core radii with \texorpdfstring{\citet{tarricq_structural_2022}}{Tarricq et al. (2022)} \label{ch:appendixB}}

\begin{figure}[ht]
    \centering
    \includegraphics[width=0.9\linewidth]{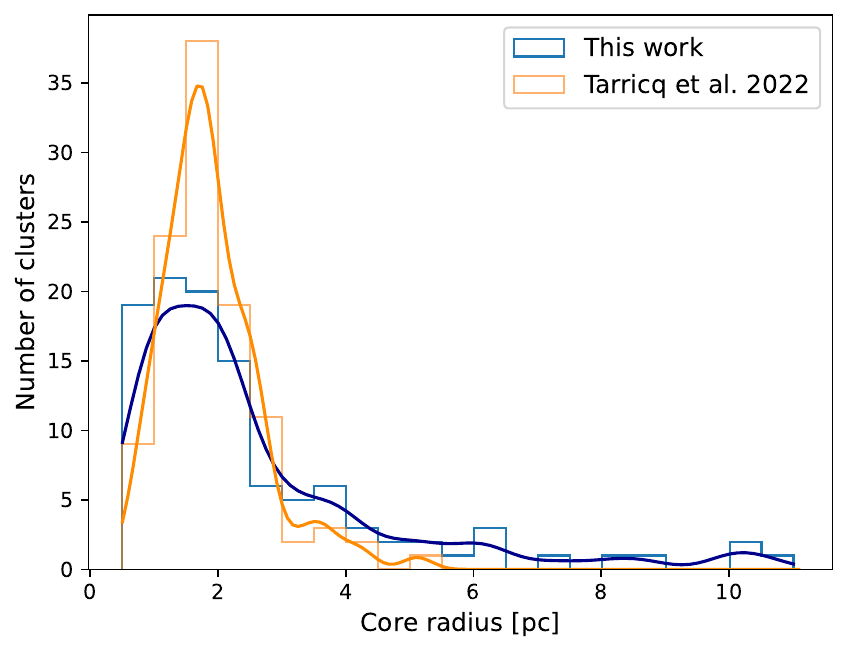}\
    \includegraphics[width=0.9\linewidth]{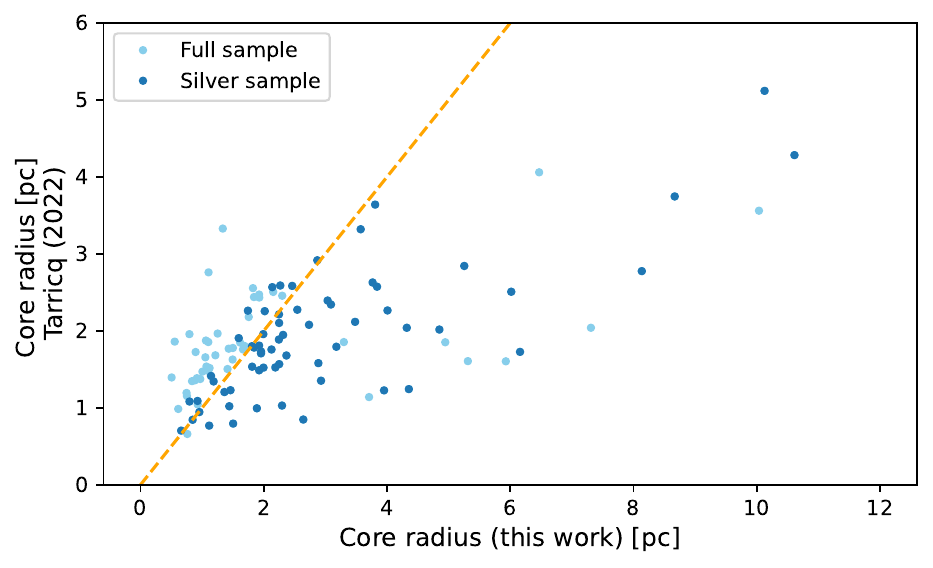}
    \caption{(Top) Distribution of core radii from this work (blue) and from \citet{tarricq_structural_2022} (orange) with fitted Gaussian KDEs. (Bottom) Individual comparison of the core radii, for the 109 OCs in common with the full sample (dark blue) and 63 in common with the silver sample (light blue). The 1:1 ratio line is plotted as the orange dashed line. }
    \label{fig:tar_rc}
\end{figure}

\newpage

\section{Comparison with results using the \citet{Chabrier_2003} IMF \label{ch:appendixC}}

In this appendix, we assess the systematic differences in our results that would arise from adopting the \citet{Chabrier_2003} IMF instead of the \citet{kroupa_2001} IMF. To this end, we replicated the mass determination procedure described in Sect.~\ref{ch:mass_det}, selecting the \citet{Chabrier_2003} IMF in the PARSEC web interface.

\begin{figure}[ht]
    \centering
    \includegraphics[width=0.9\linewidth]{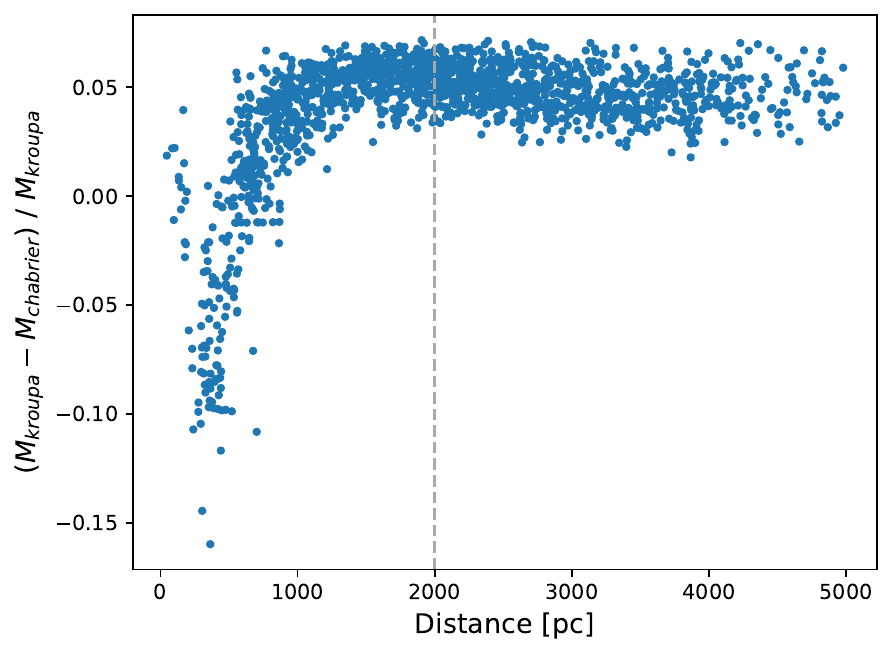}
    \caption{Fractional mass difference between the photometric masses determined using luminosity functions sampled from the \citet{kroupa_2001} and \citet{Chabrier_2003} IMFs, plotted as a function of distance. The dashed line indicates the distance limit of the sample used in our model.}
    \label{fig:mass_chabrier_kroupa}
\end{figure}

\begin{figure}[ht]
    \centering
    \includegraphics[width=0.86\linewidth]{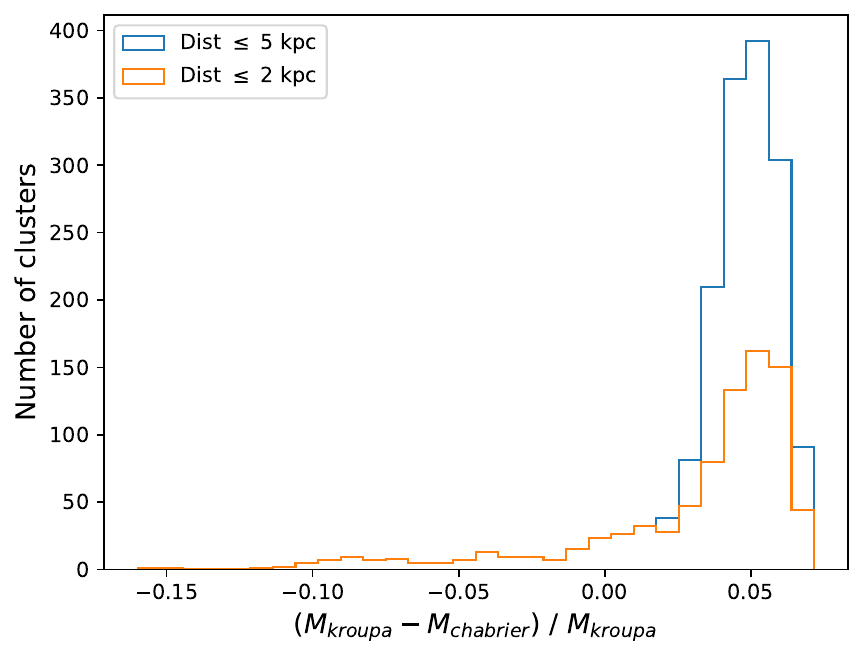}
    \caption{Distribution of the fractional mass difference for a sample cut at 5 kpc (blue) and at 2 kpc (orange).}
    \label{fig:mass_chabrier_kroupa_2}
\end{figure}

Figure~\ref{fig:mass_chabrier_kroupa} plots the fractional difference in the mass determinations (Kroupa minus Chabrier) as a function of distance. It shows that, except for a few isolated points, the difference systematically decreases from about $-10\%$ to $4\%$ for distances under approximately 1 kpc, and remains at around $4\%$ beyond that. This distance dependence arises because at closer distances, low-mass stars are included in the luminosity functions, enhancing the regime where the differences between the Chabrier and Kroupa IMFs are most pronounced. The distribution of the fractional mass differences is presented in Fig.~\ref{fig:mass_chabrier_kroupa_2}.

We note that the differences observed here were determined from luminosity function fits that assume no selective mass loss due to dynamical evolution. Therefore, there is a systematic effect due to dynamical evolution that is not being quantified in this analysis.

Since the objective of determining masses was to use them for estimating the disruption timescale, $t_4$, we reran our model using the masses derived with the Chabrier IMF. We obtained very similar results for the disruption parameters, with the optimal values being $t_4^{\text{tot}} = 2.8 \pm 0.4$ Gyr and $\gamma = 0.68 \pm 0.03$, compared to $t_4^{\text{tot}} = 2.9 \pm 0.4$ Gyr and $\gamma = 0.70 \pm 0.03$ obtained with the Kroupa IMF.

\end{appendix}

\end{document}